\documentclass[12pt]{article}

\renewcommand{\title}[1]{
\begin{center} \Large \bf #1 \end{center}
}

\renewcommand{\author}[2]{
 \begin{center} #1  \vspace{3mm} \\
  #2 \\
 \end{center}
\addvspace{\baselineskip}
}

\usepackage{amssymb}
\usepackage{amsmath}

\usepackage{amsthm}
\newtheorem{thm}{Theorem}[section]

\newtheorem{lem}[thm]{Lemma}
\newtheorem{pf}{Proof}

\theoremstyle{definition}

\theoremstyle{remark}

\makeatletter
\@addtoreset{equation}{section}

\makeatother
\newcommand{\Dirac}{{\cal D}_A}
\newcommand{\Diracb}{{\bar{\cal D}}_A}

\begin{document}

\baselineskip 5mm

\title{Noncommutative Deformation of Spinor Zero Mode and ADHM Construction}

\author{${}^{\flat}$Yoshiaki Maeda, ${}^\dagger$Akifumi Sako }{
${}^{\flat}$ Department of Mathematics,
Faculty of Science and
 Technology, Keio University\\
3-14-1 Hiyoshi, Kohoku-ku, Yokohama 223-8522, Japan \\
${}^{\flat}$ Mathematical Research Centre, 
University of Warwick\\
Coventry, CV4 7AL , United Kingdom \\
${}^\dagger$ 
Department of General Education, Kushiro National College of Technology\\
Otanoshike-Nishi 2-32-1, Kushiro 084-0916, Japan }
%{ maeda@math.sci.keio.ac.jp, ${}^\dagger$sako@kushiro-ct.ac.jp }

\vspace{1cm}

\abstract{
A method to
construct noncommutative instantons
as deformations 
from commutative instantons was provided in \cite{maeda_sako2}.
Using this noncommutative deformed instanton, 
we investigate the spinor zero modes of 
the Dirac operator in a
noncommutative instanton background 
on noncommutative ${\mathbb R}^4$, and 
%%%%@
%we show that the index of the Dirac operator is preserved
%under the noncommutative deformation.
we modify the index of the Dirac operator 
on the noncommutative space slightly and show that
the number of the zero mode of the Dirac operator is preserved
under the noncommutative deformation.
%%%%%%@
We prove the existence of the Green's function 
associated with
instantons on noncommutative ${\mathbb R}^4$,
as a smooth deformation of the commutative case.
The feature of the zero modes of the Dirac operator and 
the Green's function 
derives noncommutative 
ADHM equations which coincide with the ones introduced by
Nekrasov and Schwarz.
We show a one-to-one correspondence between
the instantons on noncommutative ${\mathbb R}^4$ and ADHM data.
An example of a noncommutative instanton
and a spinor zero mode are also given.                 
}

%%%%%%%%%%%%%%%%%%%%%%%%%%%%%%%%%%%%%%%%%%%%%%%%%%%%%%%%%%%%%%%%
\section{Introduction}
\label{section1}
%%%%%%%%%%%%%%%%%%%%%%%%%%%%%%%%%%%%%%%%%%%%%%%%%%%%%%%%%%%%%%%%%%%%%%%%%%%

%%%%@
Deformation Quantization, introduced by Flato et al \cite{Bayen:1977ha}
provided an idea for the method of
quantization, whose crucial point is not to employ the 
representation space, and treat it from purely algebraic point of view.  
It might be worth to apply this idea to gauge theories for geometry
and physics. 
We attempt to study noncommutative instantons 
in the context of the deformation 
quantization, which is one of the important problems in gauge theories.
%% One of the important problems for 
%% the gauge theory is an instanton, which has been tried 
%% by several people. 
%% Our approach presented in this paper is
%% to develop it in the context of the deformation quantization.

%We consider the gauge theory 
%There are many approaches for noncommutative gauge theories.
%One of the popular way to give them is using Moyal product
%(see Section \ref{sect2}).
%For physical reason, it is natural to restrict considering fields
%as $L^2$ functions or Schwartz functions.
%However, we do not know how can we construct the such function spaces
%to be closed under the Moyal product of two fields???.
%Therefore it is more suitable to study qualitative nature of noncommutative 
%gauge theory
%%%@

Nekrasov and Schwarz \cite{NCinstanton} discovered 
noncommutative ADHM equations and constructed 
noncommutative instantons using the ADHM construction
\cite{ADHM}. (In the following, we call these solutions briefly
 noncommutative ADHM
instantons.) 
This work initiated the study of noncommutative ADHM instantons,
and at present there is a large body of work on this problem
\cite{NCinstlecture}.
Several noncommutative instantons 
have been discovered, by the ADHM method. However, 
some of them like $U(1)$ instantons do not smoothly connect to
commutative instantons.
\footnote{ There are several noncommutative instanton solutions 
whose commutative limits have been studied,
and some of them are constructed
without the ADHM method \cite{CommLimi}.}
In \cite{sako2,sako3,Furuuchi1,Furuuchi2,Tian},
the topological charge of a noncommutative ADHM instanton is studied,
where it is shown that the topological charge 
is given by an integer and coincides with the dimension of a vector
space appearing in the ADHM construction.
(Strictly speaking, the proof of the equivalence between
the topological charge defined as the 
integral of the second Chern class and the instanton number 
given by the dimension of the vector space in the ADHM construction
is not completed. In \cite{sako3}, this identification is shown when
the noncommutative parameter is self-dual for a $U(N)$ gauge theory.
In \cite{Tian}, the equivalence is shown with
no restrictions on the noncommutative parameters, but 
a noncommutative version of the Osborn's
identity (Corrigan's identity) is assumed.)
However, the relation between the topological number
and the corresponding numbers in the commutative space had not be clarified. 
Moreover, the calculation in \cite{sako2,sako3} shows that
the origin of the instanton number is deeply related to the
noncommutativity.

On the other hand, we have constructed previously 
new noncommutative deformations of
solitons in gauge theories. These deformations smoothly connect
a commutative soliton to a noncommutative soliton
\cite{maeda_sako2,maeda_sako,sako4}.
In the following, we call these smooth noncommutative deformed
instantons SNCD instantons for short.
The SNCD instantons have a formal power series expansion in the
noncommutative parameter, and the leading terms are instantons in 
commutative space.
In particular, this produces instanton solutions 
on noncommutative ${\mathbb R}^4$
which are deformations of instanton solutions on commutative
${\mathbb R}^4$. 
We showed that the instanton
numbers of these noncommutative instanton solutions coincide
with the commutative instanton numbers on ${\mathbb R}^4$.
Thus, it is natural to ask if there is a correspondence between 
the SNCD instantons and the noncommutative ADHM construction.
Answering this question is the main purpose of this paper.

%%%%%%%%%%%%%% 27 Mar.2010
In the section 3.3 in \cite{NCinstanton}, 
the completeness of the noncommutative ADHM construction 
is already discussed without any proofs.
In this paper, we give a complete proof for the completeness
for the SNCD instanton. The procedure of the proof is basically
followed by the commutative case. One of the crucial points for the
differences to the commutative case is to observe the decay 
properties for the instantons on the
noncommutative 4-space which has been obtained in \cite{maeda_sako2}.  
Moreover, we have to have the asymptotic behavior of the 
zero modes of the Dirac operator associated with the SNCD instanton,
the index of the Dirac operator, and 
the Green's function with the SNCD instanton background. 
%%%%%%%%%%%%%%
In this article, we first investigate zero modes of the
Dirac operator associated with the SNCD instanton.
We give a (modified) index of the Dirac operator on the noncommutative space.
It is shown that this index is determined by the index 
associated with the commutative instanton.
We show the existence of the Green's function with the 
background SNCD instanton.
Using these properties, we derive the noncommutative ADHM equations 
from the SNCD instanton.
The ADHM equations coincide with the
ones discovered by Nekrasov and Schwarz \cite{NCinstanton,NCinstlecture}.
We construct one example of a SNCD instanton 
deformed from a $k=1$ BPST
instanton in commutative ${\mathbb R}^4$, 
and we check its consistency with the theorems in this
article.
In the Appendix, we show that there is one-to-one 
correspondence between the ADHM data and the 
SNCD instantons.

This paper is organized as follows. In Section 2, we set the 
notation and review basic facts about star products and 
SNCD instantons. In Section 3, we 
show that the (modified) index of the Dirac operator is constant under 
noncommutative deformations. In Section 4, we construct the Green's 
function for the noncommutative Laplacian. In Section 5, we 
prove the main result, that the ADHM equations derived from 
noncommutative instantons are the same 
as the equations constructed by Nekrasov and Schwarz. In Section 6, 
we give a worked example of a noncommutative instanton. Section 7 is 
the conclusion. 
In Appendix \ref{spinor_completeness}, some extension of the completeness
relation of the Dirac zero modes is derived.
In Appendix \ref{appendix}, we show the one-to-one correspondence 
between ADHM data and noncommutative instantons, and 
in Appendix \ref{appendixB}, we 
discuss constraints imposed by the choice of the $U(N)$ gauge group.

%%%%%%%%%%%%%%%%%%%%%%%%%%%%%%%%%%%%%%%%%%%%%%%%%%%%%%%%%%%%%%%%%%%%%%%%%%%
%%%%%%%%%%%%%%%%%%%%%%%%%%%%%%%%%%%%%%%%%%%%%%%%%%%%%%%%%%%%%%%%%%%%%%%%%%%

%%%%%%%%%%%%%%%%%%%%%%%%
\section{ Notations, Definitions and Known Facts} \label{sect2}
%%%%%%%%%%%%%%%%%%%%%%%
Noncommutative Euclidean 4-space ${\mathbb R}^4$ is given by the following commutation 
relations of the coordinates:
\begin{eqnarray}
[ x^{\mu} , x^{\nu} ]_\star = 
x^{\mu} \star x^{\nu} - x^{\nu} \star x^{\mu}= i \theta^{\mu \nu} ,
\ \mu , \nu = 1,2,3,4 \ ,
\end{eqnarray}
where $(\theta^{\mu \nu})$ is a real, $x$-independent, skew-symmetric matrix, 
whose entries are called the
noncommutative parameters.
$\star$ is known as 
the Moyal  (or star) product \cite{Moyal}.
%%%%%%%%%%%% rev1
%%%%%%%%%%%%%%%%%%%%%%%%%%
To consider smooth noncommutative deformations,
we introduce a parameter $\hbar$ and a fixed real constant
$-\infty <\theta^{\mu \nu}_0 < \infty$ with
\begin{eqnarray}
\theta^{\mu \nu} = \hbar \theta^{\mu \nu}_0 .
\end{eqnarray}
We define the commutative limit by letting $\hbar \rightarrow 0$.
%%%%%%%%%%%%%%%%%%%%%%%%%%

The Moyal product is defined on functions by
\begin{eqnarray}
 f(x)\star g(x)
  &:=&f(x)\exp\left(\frac{i}{2}\overleftarrow{\partial}_{\mu}
	      \theta^{\mu\nu}\overrightarrow{\partial}_{\nu}\right)g(x)
  \nonumber\\
 &=&f(x)g(x)+\sum_{n=1}^{\infty}\frac{1}{n!}f(x)
  \left(\frac{i}{2}\overleftarrow{\partial}_{\mu}
 \theta^{\mu\nu}\overrightarrow{\partial}_{\nu}\right)^ng(x)\;.
 \nonumber
\end{eqnarray}
%%%%! extra \epsilon in this formula?
Here $\overleftarrow{\partial}_{\mu}$ and $\overrightarrow{\partial}_{\nu}$ are  partial derivatives with
respect to $x^{\mu}$ for $f(x)$ and 
to $x^{\nu}$ for $g(x)$, respectively.

%%%%%%% Rev-2-1
%%%%%%% rev1
%%% old ver %%%We expand all fields in $\hbar$ as 
Moreover, we consider $\hbar$-expansions 
functions $f(x)$ (formal power series in $\hbar$ 
with the values in $C^{\infty} ({\mathbb R}^4)  $) 
in the following:
\begin{eqnarray}
f(x) = \sum_{n=0}^{\infty} f^{(n)}(x) \hbar^n , \label{formalexpand}
\end{eqnarray}
where $f^{(n)}(x) \in C^{\infty} ({\mathbb R}^4) $.
We mainly consider each
$f^{(n)}(x) \in C^{\infty} ({\mathbb R}^4)\cap L^2({\mathbb R}^4)$.
%%%%%%%%%%
%\begin{eqnarray}
% f^{(n)} (x)= \left(\frac{{\partial}}{\partial \hbar} \right)^n
%f(x) \Big|_{\hbar = 0} .
%\end{eqnarray}
%%
%%%%%%% Rev-2-2
%%%%%%%% rev2
We extend the Moyal product to the above fields (\ref{formalexpand})
and also to other fields like spinors $\hbar$ linealy.
%%%%%%% 
In the following, we consider all subjects by using this formal expansion 
and solve equations 
%%% Rev-2-3
(Dirac equations, etc.)
%%%
recursively in increasing orders of $\hbar$.
%%%%%%%%% Rev-2-3 %%%% move to section 3
%For example, as in Section \ref{section_Index} we treat the index theorem
%in a formal deformation setting. The index is defined by the 
%difference between the kernel and the cokernel of the formal 
%Dirac operator acting on the formal expansions.
%Also, introducing a ($\hbar$-valued ) 
%inner product in the formal expansions, 
%we employ the orthonormal bases as in 
%(\ref{3_18}), where 
%$(C^{\infty}({\mathbb R}^4) \cap L^2({\mathbb R}^4))[[\hbar ]]$
%is considered only as formal expansion space.
%%%%%%%%%%%%

We often use order estimates in the radius $|x|$. 
If $s$ is a function on ${\mathbb R}^4$
and $s=O(|x|^{-m})$, the ``natural growth condition"
is defined by $|\partial_{\mu}^k s|=O(|x|^{-m-k})$.
In this article, this natural growth condition
of gauge fields and spinor fields is 
always required.
\footnote{$s = O'(|x|^{-m})$ is defined by $s=O(|x|^{-m})$ 
and $|\partial_{\mu}^k s|=O(|x|^{-m-k})$ in \cite{D-K}. 
We do not use this symbol $O'(|x|^{-m})$ 
because it is not standard in physics.
}

%$\epsilon^{\mu \nu}$ is an anti-symmetric tensor.
%$\epsilon^{1 2} =\epsilon^{1 3}=\epsilon^{1 4}=\epsilon^{3 4}=  1$.

We define a Lie algebra structure by
$
[ {\bf T}_a , {\bf T}_b ] =  f_{abc} {\bf T}_c ,
$
where the generators ${\bf T}_a$ are anti-Hermitian matrices.
In this article, $U(N) \ (N>1)$ gauge theory on noncommutative $\mathbb{R}^4$ 
is considered. 
%The definition of our gauge transformation group action
%is given by the Moyal project, for example 
%$g^{\dagger} \star g = I_{n\times n}$
%, where $g$ is an element of 
%gauge transformation group $G$.
%Its expression in the formal expansion is seen in \cite{maeda_sako}.
The covariant derivative for a some fundamental representation 
field $f(x)$
is defined by 
\begin{eqnarray}
D_{\mu} \star f(x) := \partial_{\mu}f(x) +  A_{\mu} \star f(x) \ , \ \ 
A_{\mu}= A_{\mu}^a {\bf T}_a .
\end{eqnarray}
A gauge transformation of $A$ is given by
$A \rightarrow A + g \star d g^{-1}$
, where $g$ is an element of the
gauge group $G=\{ g\ |\ g^{\dagger} \star g = I_{n\times n} \} $.
%The definition of our gauge transformation group action
%is also given by the Moyal project, for example 
%$g^{\dagger} \star g = I_{n\times n}$.
Here $g$ has a formal expansion 
$\displaystyle g=\sum_{l=0}^{\infty} g^{(l)} \hbar^l $.
%We expand all fields with the $\hbar$. 
%$g$ is expanded as $g= \sum_{l=0}^{\infty} g^{(l)} \hbar^l$.
As we see in \cite{maeda_sako} and Appendix \ref{appendixB}, 
$g^{\dagger} \star g = I$ is equivalent to an
infinite hierarchy of algebraic equations
which we can solve recursively starting with the $\hbar^0$ term.
%Therefore, at least a formal expansion of $g$ exist.
The Laplacian is defined by
\begin{eqnarray}
\Delta_A \star f := D^{\mu} \star D_{\mu} \star f .
\end{eqnarray}
The curvature two-form $F$ is defined by
\begin{eqnarray}
F := \frac{1}{2} F_{\mu \nu} dx^{\mu} \wedge  dx^{\nu} 
= dA+  A \wedge \star A ,
\end{eqnarray}
where $\wedge \star $ is defined by 
\begin{eqnarray}
A\wedge \star A := \frac{1}{2}
(A_{\mu} \star A_{\nu} ) dx^{\mu} \wedge dx^{\nu}  .
\end{eqnarray}
%%%%%%%%%%%%%%%%%%%%%%%%%%%%%
Let ${\cal S} = {\cal S}^+ \oplus {\cal S}^-$ be the spinor bundle of ${\mathbb R}^4$.
We define $\sigma_{\mu}$ and $\bar{\sigma}_{\mu}$ by
\begin{eqnarray}
(\sigma_{1}, \sigma_{2},
\sigma_{3}, \sigma_{4})  
:= (- i {\tau}_1, - i \tau_2,- i \tau_3, I_{2\times 2} ) 
, \nonumber \\ 
(\bar{\sigma}_{1}, \bar{\sigma}_{2},
\bar{\sigma}_{3}, \bar{\sigma}_{4}) := ( i \tau_1,  i \tau_2, i \tau_3, I_{2\times 2} ) ,
\end{eqnarray} 
where $\tau_i$ are the Pauli matrices:
\begin{eqnarray}
\tau_1= \left( 
\begin{array}{cc}
0 &\ 1 \\
1 &\ 0
\end{array}
\right), \
\tau_2= \left( 
\begin{array}{cc}
0 & -i \\
i & 0
\end{array}
\right), \
\tau_3= \left( 
\begin{array}{cc}
1 & 0 \\
0 & -1
\end{array}
\right), \
\end{eqnarray}
 and $I_{2\times 2}$ is the identity matrix
of dimension 2. Note that $\sigma^{\dagger}_{\mu} = \bar{\sigma}_{\mu}$.
$\sigma_{\mu}$ and $\bar{\sigma}_{\mu}$ are a 
2-dimensional matrix representation
of the quaternions such, i.e.
\begin{eqnarray}
\sigma_{\mu}  \bar{\sigma}_{\nu} + 
\sigma_{\nu}  \bar{\sigma}_{\mu} 
= \bar{\sigma}_{\mu}   \sigma_{\nu} +
\bar{\sigma}_{\nu}   \sigma_{\mu}
 = 2\delta_{\mu \nu} .
\end{eqnarray}
We define $\sigma_{\mu \nu}$ and $\bar{\sigma}_{\mu \nu}$
as
\begin{eqnarray}
\sigma_{\mu \nu} := \frac{1}{4} (\sigma_{\mu} \bar{\sigma}_{\nu}
-\sigma_{\nu} \bar{\sigma}_{\mu} ), \  
\bar{\sigma}_{\mu \nu}:= \frac{1}{4} (\bar{\sigma}_{\mu} {\sigma}_{\nu}
-\bar{\sigma}_{\nu} {\sigma}_{\mu} ),
\end{eqnarray}
which are the components anti-selfdual and selfdual 
two-form,
%for the suffix $ \mu \nu$
respectively.
%%%%%%%%% rev6
The Dirac(-Weyl) operators 
$\Dirac \star : \Gamma ({\cal S}^{+}\otimes E) [[\hbar ]]\rightarrow
\Gamma ({\cal S}^{-} \otimes E) [[\hbar ]]$ and 
${\Diracb} \star : \Gamma ({\cal S}^{-}\otimes E)[[\hbar ]] \rightarrow 
\Gamma ({\cal S}^{+} \otimes E) [[\hbar ]]$ are defined by
\begin{eqnarray} \label{2_12}
\Dirac \star := \sigma^{\mu} D_{\mu} \star \ \mbox{and} \
\Diracb \star:= \bar{\sigma}^{\mu} D^{\dagger}_{\mu} \star \ ,
\end{eqnarray}
respectively.
%%%%%%%% rev6 

Instanton solutions or anti-selfdual
connections  satisfy the (noncommutative) instanton equation
\begin{eqnarray}
F^+ = \frac{1}{2} (1 + *) F = 0 \ , \label{ASDEQ}
\end{eqnarray}
where $*$ is the Hodge star operator.
Note that in this article instantons are anti-selfdual connections, 
not selfdual connections.
Formally, we expand the connection as
\begin{eqnarray}
A_{\mu} = \sum_{l=0}^{\infty} A_{\mu}^{(l)} \hbar^{l}  .
\end{eqnarray}
Then,
\begin{eqnarray}
A_{\mu} \star A_{\nu} &=& \sum_{l,m,n=0}^{\infty}
\hbar^{l+m+n} \frac{1}{l~ !} A_{\mu}^{(m)} 
(\overleftrightarrow{\Delta} )^l A_{\mu}^{(n)} ,
\end{eqnarray}
where
\begin{eqnarray}
\overleftrightarrow{\Delta} &\equiv& \frac{i}{2} \overleftarrow{\partial}_{\mu}
\theta^{\mu \nu}_0
\overrightarrow{\partial}_{\nu} . \nonumber
\end{eqnarray}
Using the selfdual projection operator %$P$ as
\begin{eqnarray}
P := \frac{1+*}{2}  \ ; \ 
P_{\mu \nu , \rho \tau} 
=\frac{1}{4} (\delta_{\mu \rho} \delta_{\nu \tau} -
\delta_{\nu \rho} \delta_{\mu \tau} + \epsilon_{\mu \nu \rho \tau} ) ,
\label{SDProjection}
\end{eqnarray}
the instanton equation is 
\begin{eqnarray} \label{ASDEQ_P}
P_{\mu \nu , \rho \tau} F^{\rho \tau} =0 . 
\end{eqnarray}
In the noncommutative case, the $l$-th order equation of (\ref{ASDEQ_P})
is given by
\begin{eqnarray}
&&P^{\mu \nu , \rho \tau} 
( \partial_{\rho} A_{\tau}^{(l)}- \partial_{\tau} A_{\rho}^{(l)}
+ i [A_{\rho}^{(l)} , A_{\tau}^{(l)} ] + C_{\rho \tau}^{(l)} ) = 0
,  \label{k-th_order}
\end{eqnarray}
where
\begin{eqnarray}
&&C_{\rho \tau}^{(l)} := 
\sum_{(p ;~ m,n) \in I(l)} 
\hbar^{p+m+n} \frac{1}{p~ !} \big( A_{\rho}^{(m)} 
(\overleftrightarrow{\Delta} )^p A_{\tau}^{(n)} -
A_{\tau}^{(m)} 
(\overleftrightarrow{\Delta} )^p A_{\rho}^{(n)}
\big)\  , \nonumber \\
&& I(l) \equiv 
\{( p ;~ m,n ) \in {\mathbb Z}^3 | p+m+n =l,~ p,m,n \ge 0 ,~
m\neq l , n\neq l \} . \nonumber
\end{eqnarray}
Note that the zeroth order equation is the commutative instanton equation 
with solution  $A^{(0)}_{\mu}$
a commutative instanton.
The asymptotic behavior of the commutative instanton $A^{(0)}_{\mu}$
is given by
\begin{eqnarray}
A^{(0)}_{\mu} = g^{(0)} \partial_{\mu} {(g^{(0)})}^{-1}+O(|x|^{-2}), \ \ 
g^{(0)}d {(g^{(0)})}^{-1} = O(|x|^{-1}) ,
\label{AsymptoticA}
\end{eqnarray}
where $g^{(0)}$, the zeroth order term in the expansion of $g$,
is an element of the gauge group in commutative
space. 
We impose a boundary condition that is a natural extension of 
(\ref{AsymptoticA}):
\begin{eqnarray}
A_{\mu} = g \star \partial_{\mu} g^{-1}+O(|x|^{-2}), \ \ 
g \star d g^{-1} = O(|x|^{-1}) .
\label{AsymptoticB}
\end{eqnarray}
In \cite{maeda_sako2}, we found a solution of (\ref{k-th_order}),
which we call a SNCD instanton.
The order of the SNCD instanton is given by
\begin{eqnarray}
A^{(l)}_{\mu} = O(|x|^{-3+\epsilon}) \ , l=1,2,3, \dots ,  
\label{3epsilon}
\end{eqnarray}
for arbitrarily small $\epsilon >0 $.
We denote (\ref{3epsilon}) by $A^{(l)} = O(|x|^{-3+\epsilon})$ for simplicity.
We proved also that the instanton number of SNCD coincides with the
instanton number of $A^{(0)}$:
\begin{eqnarray}
\frac{1}{8\pi^2} \int tr F \wedge \star F
=\frac{1}{8\pi^2} \int tr F^{(0)} \wedge  F^{(0)} .
\end{eqnarray}

%%%%
For a later convenience, we introduce covariant derivatives associated to the commutative instanton
connection by
\begin{eqnarray}
D^{(0)}_{\mu} f := \partial_{\mu} f +  A_{\mu}^{(0)}  f  ,\ \
\end{eqnarray}
%Using this, (\ref{k-th_order}) is given by
%%%%! either of the equivalent equations (?)
%\begin{eqnarray}
%&&P^{\mu \nu , \rho \tau} \big( 
%D^{(0)}_{\rho} A_{\tau}^{(l)}- D^{(0)}_{\tau} A_{\rho}^{(l)}
%+ C_{\rho \tau}^{(l)}
%\big) =0 \nonumber \\
%&& P (D_{A^{(0)}} A^{(l)} + C^{(l)} )=0 . \label{lthASD}
%\end{eqnarray}
and the Laplacian associated with the commutative instanton
connection by
\begin{eqnarray}
\Delta_A^{(0)} f := D^{(0) \mu} D^{(0)}_{\mu} f  .
\end{eqnarray}

%%%%%%% rev3 %%%%%%%%%%%%%
Let us introduce a $\hbar$-valued 
%%%%%% Rev-2-4
%global inner product 
pairing
for formal expansions $f(x), g(x) \in (C^{\infty}({\mathbb R}^4) \cap L^2({\mathbb R}^4))[[\hbar ]]$ as
\begin{eqnarray} \label{formal_inner_product}
\langle f , g \rangle_{\star} 
:= \int_{{\mathbb R}^4} d^4 x ( f^{\dagger}(x) ,  g(x) )_\star .
%=\int_{{\mathbb R}^4} d^4 x (f(x)^{\dagger} , g(x)).
\end{eqnarray}
Here $ (\ \ ,\ \  )_{\star} $ is the $\hbar$-valued 
point wise product used in Euclidean scalar product 
with contraction of spinors or tensors, that is  
%rotation group $SO(4)$ (or $SU(2)\times SU(2)$) and 
$( f^{\dagger}(x) ,  g(x) )_\star$ is defined by
\begin{eqnarray}
( f^{\dagger}(x) ,  g(x) )_\star:= f^{\dagger \mu_1 , \dots \mu_n } \star 
g_{\mu_1 \dots \mu_n } .
\end{eqnarray}
Since each $f^{(n)}$ and $g^{(n)}$ are in $C^{\infty}({\mathbb R}^4) \cap L^2({\mathbb R}^4)$,
we obtain
\begin{eqnarray}
\int_{{\mathbb R}^4} d^4 x f^{(n)}(x) \overleftrightarrow{\Delta} g^{(m)}(x)
=0 .
\end{eqnarray}
then 
\begin{eqnarray} 
\langle f , g  \rangle_{\star} 
%:= \int_{{\mathbb R}^4} d^4 x ( f^{\dagger}(x) ,  g(x) )_\star .
&=& \int_{{\mathbb R}^4} d^4 x (f(x)^{\dagger} , g(x)) \nonumber \\
&=& \sum_{n=0}^{\infty}
\int_{{\mathbb R}^4} d^4 x  \sum_{k+l=n} (f(x)^{\dagger (k)} , g^{(l)}(x))
\hbar^n
,
\end{eqnarray}
where
$( f^{\dagger}(x) ,  g(x) )$ is defined by
$
( f^{\dagger}(x) ,  g(x) ):= f^{\dagger \mu_1 , \dots \mu_n } 
g_{\mu_1 \dots \mu_n } $.
%It seems reasonable to think this pairing (\ref{formal_inner_product})
%as a global inner product for formal expansions $f(x)$ and $g(x)$.
We also use the usual $L^2$ inner product, that is for $\hbar$
independent function $f(x), g(x)$, we set
\begin{eqnarray}
\langle f(x) , g(x) \rangle := \int _{{\mathbb R}^4} d^4 x 
f^{\dagger} (x) g (x) .
\end{eqnarray}
If $f(x)$ and $g(x)$ are not scalar functions, 
we regard $f^{\dagger} (x) g (x)$ as a point wise production with contraction.

We note that 
our formal space 
$(C^{\infty}({\mathbb R}^4) \cap L^2({\mathbb R}^4))[[\hbar ]]$
is considered only as a formal expansion space.
%%%%%%%% rev4 %%%%%%%%
%%%%%%%%%%%%%%%%%%%%%%%%%%

%%%%%%%%%%%%%%%%%%%%%%%%%%%%%%%%%%%%%%%%%%%%%%%
%%%%%%%%%%%%%%%%%%%%%%%%%%%%%%%%%%%%%%%%%
%%%%%%%%%%%%%%%%%%%%%%%%%%%%%%%%%%%%%%%%%%
\section{The Index of the Dirac Operator} \label{section_Index}
In this section, we investigate zero modes of the Dirac operators
acting on the formal expansion space.
The index theorem for the Dirac operator in a 
noncommutative ADHM instanton background was studied in \cite{Kim:2002qm},
where it was shown that the number of zero modes of the Dirac operator
equals the instanton number of the background instanton
in the ADHM construction.
In our case, we start with a commutative instanton
and deform it into a SNCD instanton.
%%%%%%%%%%% rev5
The relation between SNCD instantons and ADHM instantons
will be clarified by using the theorem \ref{index_theo} proved
in this section.
%%%%%%%%%%% rev5
To construct the ADHM data from SNCD instantons, 
we have to investigate the spinor zero modes and the index.
In Section \ref{InstantonADHM}
these results will be used to derive
the ADHM equations.\\

%%% Rev-2-3 %%%@ from section 1
In this article, we treat the index theorem
in a formal deformation setting. The usual index is defined by the 
difference between the kernel and the cokernel of the 
Dirac operator.
In our context, the differential operators like the Dirac operator act on 
$(C^{\infty}({\mathbb R}^4) \cap L^2({\mathbb R}^4))[[\hbar ]]$
that is considered only as formal expansion space.
We introduce a ($\hbar$-valued ) 
inner product in the formal expansions, 
we employ the orthonormal bases as in 
(\ref{3_18}), and then we define the modified index as (\ref{modified_index}).
%%%%%@

We consider operators acting on the weighted Sobolev spaces 
\begin{eqnarray}
\widetilde{W}_{\delta}^{k,p} =
\Big\{ \vec{u} %= \sum_{l=0}^\infty \vec{u}_l \hbar^l  
\ \Big| \  \sum_{ j < k } || \partial^j \vec{u} ||_{p, \delta -j}  
= : || \vec{u} ||_{k,p,\delta} 
< \infty \Big\} ,
\end{eqnarray}
where $j=j_1+j_2 +j_3 +j_4$ , $ 
\partial^j=\partial^{j_1}_1 \partial^{j_2}_2 
\partial^{j_3}_3 \partial^{j_4}_4$,
and 
\begin{eqnarray}
|| \vec{u} ||_{p, \delta} := \Big| \int_{{\mathbb R}^4} 
\{ (1+|x|^2)^{\frac{1}{2}} \}^{-\delta p -4} 
|\vec{u}(x)|^p dx \Big|^{\frac{1}{p}} < \infty .
\end{eqnarray}
Here $|\vec{u}(x)| = \sqrt{\vec{u}^{\dagger} \vec{u}}$.
See \cite{Bartnik} for the properties of weighted Sobolev spaces 
used here.
%%%%%%%%% Rev-2-7
We do not introduce the norm from the pairing (\ref{formal_inner_product})
usual to complete spaces of $\hbar$ - expansions.
We deal with the Hilbert spaces (Sobolev spaces) as usual $L^2$-space 
step by step for $\hbar$ - expansions.

Let $\Dirac \star : \Gamma({\cal S}^+ \otimes E)[[ \hbar ]]
\rightarrow \Gamma({\cal S}^- \otimes E)[[ \hbar ]]
$ and
$\Diracb \star : \Gamma({\cal S}^- \otimes E)[[ \hbar ]]
\rightarrow \Gamma({\cal S}^+ \otimes E)[[ \hbar ]]
$ be the Dirac operator defined by (\ref{2_12}).
%%%%%%%%%%%%%%%
By the Weitzenbock formula, 
\begin{eqnarray}
\Diracb \star \Dirac = \Delta_A + \sigma^{+}  F^{+} \ ,
\label{Weitzenbock1} \\
\Dirac \star \Diracb = \Delta_A + \sigma^{-}  F^{-} \ ,
\end{eqnarray}
where $\sigma^{+} F^{+}= 2 \bar{\sigma}^{\mu \nu} F^+_{\mu \nu} \ , \ 
\sigma^{-} F^{-}= 2 {\sigma}^{\mu \nu} F^-_{\mu \nu} $ and 
$\Delta_A = D^{\mu} \star D_{\mu}$.
Assume that $A$ is a noncommutative
anti-selfdual connection, i.e. $F^+=0$.
We consider the $\hbar$ expansion of 
${\psi} \in \Gamma ({\cal S}^{+} \otimes E )[[\hbar ]]$:
\begin{eqnarray}
{\psi} = \sum_{n=0}^{\infty} \hbar^{n} {\psi}^{(n)} .
\end{eqnarray}
%%%%%%%%%%%% Rev-2-8
Set 
\begin{eqnarray}
Ker \Dirac \star :=&& \left\{ \  \psi \in \Gamma ({\cal S}^{+} \otimes E ) \cap
L^2 ({\cal S}^{+} \otimes E)[[\hbar ]] \right.  \nonumber\\
&& \ \ \ \ \left|\  \Dirac \star \psi = 0 \in  
\Gamma ({\cal S}^{-} \otimes E ) [[\hbar ]]\ \  \right\} .
\end{eqnarray}
As in the commutative case, we obtain the
following theorem.
\begin{thm} \label{lem_no_zeromode}
Assume that $A$ is a SNCD anti-selfdual connection.
Then
% $Ker \Dirac \star= 0$ on 
% $\{ \psi |~ \psi^{(l)} \in L^2 \ (l= 0,1,2,\dots ) \}$,
%i.e.
%$\psi = 0$ 
If $ \Dirac \star \psi = 0 $ for $\psi^{(n)} \in L^2 $,
we have $\psi^{(n)}=0$ for all $n$, 
i.e. $Ker \Dirac \star= 0$.
\end{thm}

\noindent
\begin{pf}
We show this theorem by induction.
The zeroth order term $ \Dirac \star \psi = 0 $
is $\Dirac^{(0)} \psi^{(0)} =0$, and 
this equation only has the solution $\psi^{(0)}= 0$.
We assume that the $\psi^{(k)} = 0\ ( k \le n )$.
The equation of order $n+1$ is 
\begin{eqnarray}
0&=&\hbar^{n+1} \left\{
\Dirac^{(0)} \psi^{(n+1)} +\sigma^{\rho}  A_{\rho}^{(n+1)} 
\psi^{(0)} +
\sum_{(p ;~ l,m) \in I(n+1)} 
\frac{1}{p~ !} \big(\sigma^{\rho}  A_{\rho}^{(l)} 
(\overleftrightarrow{\Delta} )^p \psi^{(m)}
\big)\ \right\} \nonumber \\
&=& \hbar^{n+1} \Dirac^{(0)} \psi^{(n+1)}  , \nonumber 
\end{eqnarray}
so $\psi^{(n+1)} = 0$.
%Therefore, we obtain  $Ker \Dirac \star=0 $.
\mbox{}\hfill$\square $
\end{pf}

We investigate the zero modes of $\Diracb \star$.
%%%%%% rev JMP 
%%%%@
Set 
\begin{eqnarray}
Ker \Diracb \star 
:=&& \left\{ \  \bar{\psi} \in \Gamma ({\cal S}^{-} \otimes E ) \cap
L^2 ({\cal S}^{-} \otimes E)[[\hbar ]] \right.  \nonumber\\
&& \ \ \ \ \left|\  \Diracb \star \bar{\psi} = 0 \in  
\Gamma ({\cal S}^{+} \otimes E ) [[\hbar ]]\ \  \right\} .
\end{eqnarray}
%%%%%@
%%%%%%%%
By expanding 
%%%%%% rev7 add [[ \hbar ]] at several parts
$\bar{\psi} \in \Gamma ({\cal S}^{-} \otimes E )[[ \hbar ]]$ 
%%%%%%%% rev7
as
\begin{eqnarray}
\bar{\psi} = \sum_{n=0}^{\infty} \hbar^{n} \bar{\psi}^{(n)} ,
%\ , \  \ \bar{\psi}^{(n)} = O'( |x|^{-(3+2n)} ) 
\end{eqnarray}	
the zeroth order equation of $ \Diracb \star \bar{\psi} = 0 $
is $\Diracb^{(0)} \bar{\psi}^{(0)} =0$, and 
there are $k$ linearly independent 
zero-modes for a commutative instanton $A^{(0)}$
whose instanton number is $-k$.
We define $\bar{\psi}_i \ (i=1, \dots ,k )$ as 
\begin{eqnarray}
\bar{\psi}_i = \sum_{n=0}^{\infty} \hbar^n \bar{\psi}^{(n)}_i , 
\end{eqnarray}
where $\bar{\psi}^{(0)}_i (i=1, \dots ,k )$ are a basis of the $k$ 
independent zero modes of $\Diracb^{(0)}$. 

The $n$-th order equation of $ \Diracb \star \bar{\psi} = 0 $ is 
\begin{eqnarray}
0&=&\hbar^{n} \left\{
\Diracb^{(0)} \bar{\psi}^{(n)}_i +\bar{\sigma}^{\rho}  A_{\rho}^{(n)} 
\bar{\psi}^{(0)}_i +
\sum_{(p ;~ l,m) \in I(n)} 
\frac{1}{p~ !} \big(\bar{\sigma}^{\rho}  A_{\rho}^{(l)} 
(\overleftrightarrow{\Delta} )^p \bar{\psi}^{(m)}_i
\big)\ \right\} \nonumber \\
&=&
\hbar^{n} \left\{
\Diracb^{(0)} \bar{\psi}^{(n)}_i + H^{(n)}_i \right\}, 
\label{spinor_zeromode_eq}
\end{eqnarray}
where $H^{(0)}_i =0$ and 
\begin{eqnarray}
H^{(n)}_i= \bar{\sigma}^{\rho}  A_{\rho}^{(n)} 
\bar{\psi}^{(0)}_i +
\sum_{(p ;~ l,m) \in I(n)} 
\frac{1}{p~ !} \big(\bar{\sigma}^{\rho}  A_{\rho}^{(l)} 
(\overleftrightarrow{\Delta} )^p \bar{\psi}^{(m)}_i
\big)\ \ \ \mbox{for} \ n \in {\mathbb N}.
\end{eqnarray}
We can solve these equations recursively in the order in $\hbar$,
so $H^{(n)}_i$ is 
%regarded as a given function when we solve 
determined by Eq. (\ref{spinor_zeromode_eq}).
$\Diracb^{(0)} \bar{\psi}^{(n)}_i$ in Eq. (\ref{spinor_zeromode_eq}) 
has $k$ zero modes. 
We denote an orthonormal basis of $Ker \Diracb^{(0)}  $
by $\eta_i \ (i=1 \dots k )$.
Note that $\Dirac^{(0)} H^{(n)}_i $ is orthogonal to
%%%%% Rev-2-9
$Ker \Diracb^{(0)} $ with respect to usual $L^2$ inner product :
\begin{eqnarray}
\langle (\Dirac^{(0)}  H^{(n)}_j ) , \eta_i \rangle = 
\int_{{\mathbb R}^4} d^4 x  ( \Dirac^{(0)}  H^{(n)}_j )^{\dagger} \eta_i
=-
\langle H^{(n)}_j , \Diracb^{(0)}  \eta_i \rangle =0 .
\end{eqnarray}
Then we get 
\begin{eqnarray} \label{diraczeromode}
 \bar{\psi}^{(n)}_i = 
 \sum_{j=1}^{k} a_{n,i}^j \eta_j
-\frac{1}{\Dirac^{(0)} \Diracb^{(0)}} \Dirac^{(0)}
H^{(n)}_i ,
\end{eqnarray}
where $a_{n,i}^j$ are arbitrary constants.
Here
$\displaystyle \frac{1}{\Dirac^{(0)} \Diracb^{(0)}} $
denotes integration over $\mathbb{R}^4$ against
the Green's function of
${\Dirac^{(0)} \Diracb^{(0)}}$.
Note that the ambiguity in the $a_{n,i}^j$ are occur only in
the coefficients of the zero modes
of the commutative Dirac operator $\Diracb^{(0)}$.
$a_{n,i}^j$ is a constant matrix in general,
because the symmetries realized in matrix representations
remain after
noncommutative deformation.

We denote $K(x,y)$ by the kernel function 
of $(\Dirac^{(0)} \Diracb^{(0)} )^{-1}$.
We recall the Weitzenbock formula,
\[
\Dirac^{(0)} \Diracb^{(0)}=\Delta_A^{(0)} + \sigma^{-} F^{-(0)}. 
\]
The following  is known (cf. see \cite{Bartnik}, Theorem 1.7):
\begin{eqnarray}
K(x,y) = \frac{C}{|x-y|^2}  + O(\frac{1}{|x-y|^3}) .
\end{eqnarray}

In the following, we consider  %$\bar{\psi}$ , $H$ and so on as matrices
$\bar{\psi} :=(\bar{\psi}_1 , \dots , \bar{\psi}_k )$, 
$H :=(H_1 , \dots , H_k )$ as matrices.
%%%%%%%%%%%%%%%%%%%%%%%
%%%%%%%%%%%%%%%%%%%%%%%
\begin{thm} %[$\bar{\psi}$ zero mode] 
\label{lemma_A}
Assume that $A$ is a SNCD anti-selfdual connection.
Let $\displaystyle \bar{\psi}=(\bar{\psi}_i )=
\sum_{n=0}^{\infty} \psi^{(n)} \hbar^n$ be a zero mode of $\Diracb \star$ 
as above. 
%$\bar{\psi}^{(n)} = O(|x|^{-3})$ for $n \ge 1 $ .
Then%, each $\psi^{(n)}_i$ is obtained as
\begin{eqnarray}
\bar{\psi}^{(n)}_i &=& \sum_{j=1}^{k} a_{n,i}^j \eta_j  
-\frac{1}{\Dirac^{(0)} \Diracb^{(0)}} \Dirac^{(0)}
H^{(n)}_i ,\\
 \eta_j &=& O(|x|^{-3}) , \ \ 
\frac{1}{\Dirac^{(0)} \Diracb^{(0)}} \Dirac^{(0)}
H^{(n)}_i = O(|x|^{-5+\epsilon}),
\end{eqnarray}
and 
\begin{eqnarray}
\bar{\psi}_i=  \sum_{n=0}^{\infty} 
(\sum_{j=1}^{k} a_{n,i}^j \eta_j )\hbar^n  +O(|x|^{-5+\epsilon})
\ , \eta_j = O(|x|^{-3}) . \label{spinor_asym}
\end{eqnarray}
\end{thm}

\begin{pf}
We prove this theorem by induction.
(i) By (\ref{3epsilon}), $A^{(k)} = O(|x|^{-3+\epsilon})$, 
so we obtain
\begin{eqnarray}
H^{(1)}=\bar{\sigma}^{\rho}  A_{\rho}^{(1)} 
\bar{\psi}^{(0)} +
 \big(\bar{\sigma}^{\rho}  A_{\rho}^{(0)} 
(\overleftrightarrow{\Delta} ) \bar{\psi}^{(0)}
\big)= O(|x|^{-6+\epsilon}) .
\end{eqnarray}
{}From $K(x,y) = \frac{C}{|x-y|^2}$, we have
$\displaystyle
\frac{1}{\Dirac^{(0)} \Diracb^{(0)}} \Dirac^{(0)}
H^{(1)}= O(|x|^{-5+\epsilon}) $. 
The detailed derivation of this equation is similar to
the proof of Proposition 1 in \cite{maeda_sako2}. 
$\eta_i=O(|x|^{-3})$ is a well-known fact (see for example 
\cite{D-K,Charbonneau} and Appendix \ref{appendix}).
Using $\eta_i=O(|x|^{-3})$ and (\ref{diraczeromode}),
we obtain
\begin{eqnarray}
\bar{\psi}^{(1)} =O(|x|^{-3}).
\end{eqnarray}
(ii) Assume 
\[\bar{\psi}^{(l)}_i= \sum_{j=1}^{k} a_{l,i}^j \eta_j  +O(|x|^{-5+\epsilon}) ,\ (0 \le l \le n). \]
Then we obtain $H^{(n+1)}= O(|x|^{-6+\epsilon})$ and
$\displaystyle
\frac{1}{\Dirac^{(0)} \Diracb^{(0)}} \Dirac^{(0)}
H^{(n+1)}= O(|x|^{-5+\epsilon}) \ .
$
Therefore, $\bar{\psi}^{(n)}_i= \sum_{j=1}^{k} a_{n,i}^j \eta_j  +O(|x|^{-5+\epsilon})$.
\mbox{}\hfill$\square $
\end{pf}

%%%%% Rev-2-10
Note that this theorem implies that each 
$\bar{\psi}^{(n)} \in L^2 ({\cal S}^{-} \otimes E)$.
%%%%%

%%%%% rev for ver3_1
%%%%%@
We give a canonical
choice of zero modes of $\Diracb \star$ by introducing
a formal orthonormalization of the 
zero modes of $\Diracb \star$.
%%%%%@
%%%%%%%%%%%%%%%%%%%%%%%%%%%
Let $\bar{\psi}_0 $ be a zero mode of $\Diracb \star $.
%%%%%%%% Rev-2-9
%%%%% rev8
Formal expansion of the pairing 
$\langle \bar{\psi}^{\dagger}_0 , \bar{\psi}_0 \rangle_{\star}$
(defined by (\ref{formal_inner_product}))
is given by
%%%%%%% 
\begin{eqnarray}
\int_{{\mathbb R}^4} d^4 x \bar{\psi}_0^{\dagger} \star \bar{\psi}_0 
&=&\int_{{\mathbb R}^4} d^4 x \bar{\psi}_0^{\dagger} \bar{\psi}_0 
= \sum_{n=0}^{\infty} 
 \sum_{k+l=n} \int_{{\mathbb R}^4} d^4 x 
 \bar{\psi}_0^{(k) \dagger} \bar{\psi}_0^{(l)} \hbar^n \nonumber \\
&=& \sum_{n=0}^{\infty} [\langle \bar{\psi}_0^{\dagger} , 
\bar{\psi}_0 \rangle ]^{(n)} \hbar^n . \label{3_18}
\end{eqnarray}
Here we use the decay condition $\bar{\psi}_0 \rightarrow 0$
as $|x|\rightarrow \infty$.
The inverse of the formal power series of 
$\displaystyle \sum_{n=0}^{\infty} a^{(n)} \hbar^n $
with $a^{(0)} \neq 0$ 
is defined by $\displaystyle \sum_{n=0}^{\infty} b^{(n)} \hbar^n $, where
$\displaystyle
b^{(0)}=\frac{1}{a^{(0)}} $ and $\displaystyle
b^{(n)}= - \frac{1}{a^{(0)}} \sum_{i=0}^{n-1} a^{(n-i)} b^{(i)}.
$
Since $\langle \bar{\psi}_0^{(0)\dagger} , \bar{\psi}_0^{(0)} \rangle \neq 0$,
its formal inverse is defined by
\begin{eqnarray}
\left( \langle\bar{\psi}_0^{\dagger} , \bar{\psi}_0 \rangle_{\star} \right)^{-1}
:= 
\sum_{n=0}^{\infty} \hbar^n 
[\langle\bar{\psi}_0^{\dagger} , \bar{\psi}_0 \rangle_{\star}^{-1}]^{(n)} ,
\end{eqnarray}
where $[\langle\bar{\psi}_0^{\dagger} , \bar{\psi}_0 \rangle_{\star}^{-1}]^{(n)}$
is determined by
\begin{eqnarray}
[\langle\bar{\psi}_0^{\dagger} , \bar{\psi}_0 \rangle_{\star}^{-1}]^{(n)} =
- \frac{1}{\langle\bar{\psi}_0^{(0)\dagger} , \bar{\psi}^{(0)}_0 \rangle }
\sum_{i=0}^{n-1} [\langle\bar{\psi}_0^{\dagger} , \bar{\psi}_0 \rangle]^{(n-i)} 
 [\langle\bar{\psi}_0^{\dagger} , \bar{\psi}_0 \rangle^{-1}]^{(i)}  . 
%\frac{ [\langle\bar{\psi}^{\dagger} , \bar{\psi} \rangle]^{(n)} }{\langle\bar{\psi}^{(0)\dagger} , \bar{\psi}^{(0)} \rangle } 
\end{eqnarray}
This construction allows us to construct an orthonormalization.
Let the $2N \times k$ matrix $\bar{\psi} $ be a zero mode of $\Diracb \star $.
We set the following orthonormal condition 
\begin{eqnarray}
\int_{{\mathbb R}^4} d^4 x  \bar{\psi}^{\dagger} \star \bar{\psi} 
= I_{k\times k} . \label{3_21}
\end{eqnarray}
The $l$-th order equation in $\hbar$ for (\ref{3_21}) is
\begin{eqnarray}
\sum_{n+m=l} \int_{{\mathbb R}^4} d^4 x
\Big(\sum_{j=1}^{k}  \eta^{\dagger}_j a_{n,i}^{j \dagger} - {\cal H}^{n \dagger}_i \Big) \Big( 
 \sum_{j=1}^{k} a_{m,p}^j \eta_j - {\cal H}^m_p \Big)=
\delta_{i p} \delta_{l 0} ,
\end{eqnarray}
where
\begin{eqnarray}
{\cal H}^n_i = \frac{1}{\Dirac^{(0)} \Diracb^{(0)}} \Dirac^{(0)}
H^{(n)}_i .
\end{eqnarray}
Gram-Schmidt orthonormalization determines the constants
$a_{n,i}^j$ recursively.
%%%%%%% rev JMP 
%%%%@
We introduce a linear space that is expanded by these
formal orthonormalized zero modes
\begin{eqnarray}
{\widehat{Ker}} \Diracb \star 
:= \left\{ \bar{\psi}\ \right. &\Big| & \
 \bar{\psi}= \sum_{i=1}^k c_i \bar{\psi}_i 
, \ \psi_i \in Ker \Diracb \star \ , \bar{\psi}_i^{(0)}=\eta_i ,  \nonumber \\ 
 && \left.
\int_{{\mathbb R}^4} d^4 x  \bar{\psi}^{\dagger}_i \star \bar{\psi}_j 
= \delta_{ij} , c_i \in {\mathbb C} \right\} .
\end{eqnarray}
%%%%@
%%%%%%%%%%%%%%%%%

%%%%%%%% rev9
We recall the index for the $\displaystyle \not{\!\!D}^{0}_A$ is defined by
$$\displaystyle \makebox{\rm Ind} \not{\!\!D}^{0}_A := 
\mbox{\rm dim ker} \Dirac^{(0)} - \mbox{\rm dim ker}\Diracb^{(0)} $$
as usual.
We define the modified index for the $\not{\!\!D}_A \star$ as
\begin{align}
\widehat{\makebox{\rm Ind}} \not{\!\!D}_A \star
:= 
\mbox{\rm dim Ker} \Dirac \star - \mbox{\rm dim} \widehat{Ker} \Diracb \star .
\label{modified_index}
\end{align}

Thus we have the following theorem.
%%%%%%%% rev9
\begin{thm} \label{index_theo}
If $\makebox{\rm Ind} \not{\!\!D}^{0}_A  = -k$, then
 $\widehat{\makebox{\rm Ind}} \not{\!\!D}_A \star = -k$ .
\end{thm}

Note that this $\widehat{\makebox{\rm Ind}} \not{\!\!D}_A \star$
is not index in usual sense.
One reason is that the $\Dirac \star$ and the $\Diracb \star$
are not Fredholm operators because we consider formal power series.
Another reason is that $\widehat{Ker} \Diracb \star \neq {Ker} \Diracb \star$
($\widehat{Ker} \Diracb \star \subset {Ker} \Diracb \star$).
For example, if 
$\displaystyle \bar{\psi}=\sum_{n=0}\hbar^{n} \bar{\psi}^{(n)}$
is a zero mode of $\Diracb \star$, then
$\displaystyle  \bar{\psi}'=\sum_{n=0}\hbar^{n+k} \bar{\psi}^{(n)}$
is also a zero mode for arbitrary integer $k$.
We find that $\bar{\psi}' \in {Ker} \Diracb \star$ but 
$\bar{\psi}' \not\in \widehat{Ker} \Diracb \star$.
However, in our context,
it is a natural extension of the index of usual commutative space,
because the dimension of the $\widehat{Ker}$ is essential for the
construction of the ADHM data and the relation with the instanton number.

%%%%%%%%%%%%%%%%%%%%%%%%%%%%%%%%%%%%%%%%%%%%%%%%%%%%%%%%%%%
%%%%%%%%%%%%%%%%%%%%%%%%%%%%%%%%%%%%%%%%%%%%%%%%%%%%%%%%%%%

\section{Green's Function} \label{sectGreenFunction}
In this section, we construct the Green's function for $\Delta_A$.
The definition of the Green's function is
\begin{eqnarray}
\Delta_A \star G_A (x,y) = \delta_{\star} (x-y), \label{green1} 
\end{eqnarray}
where
\begin{eqnarray}
\int d^4x  \delta_{\star} (x-y) \star f(y) =f(x) .
\end{eqnarray}
Note that if $f(x)$ is smooth,
\begin{eqnarray}
\int d^4x  \delta_{\star} (x-y) \star f(y)
=  \int d^4x  \delta_{\star} (x-y) f(y) .
\end{eqnarray}
Then, 
we do not distinguish $
\delta_{\star} (x-y) $ and $ \delta (x-y)$
in the following.
%%%%%%%%%%%%% footnote
\footnote{
This discussion might be too naive.
To avoid any risk of error, we should define
$G_A (x,y)$ by 
$\Delta_A \star G_A (x,y) = \delta (x-y)$.
}

%%%%%%%%%%%%%%

We expand (\ref{green1}) in $\hbar$:
\begin{eqnarray}
\hbar^0 &:& \Delta_A^{(0)} G_A^{(0)} (x,y) =  \delta (x-y) \label{g_0}\\
\hbar^1 &:& \Delta_A^{(0)} G_A^{(1)} (x,y) + [ \Delta_A \star G_A^{(0)} (x,y) ]^{(1)} =0 \\
& \vdots &  \nonumber \\
\hbar^n &:& \Delta_A^{(0)} G_A^{(n)} (x,y) + 
[ \Delta_A \star \sum_{0\le k <n} \hbar^k G_A^{(k)} (x,y) ]^{(n)} =0  .
\label{g_n}\\
&\vdots& \nonumber
\end{eqnarray}
We solve (\ref{g_0})-(\ref{g_n}) recursively as
\begin{eqnarray}
G_A^{(n)} (x,y) = \int d^4 w G_A^{(0)} (x,w) 
[ \Delta_A \star \sum_{0\le k <n} \hbar^k G_A^{(k)} (w,y) ]^{(n)} \ .
\end{eqnarray}
Note that $G_A^{(0)} (x,w)$ was constructed in
\cite{Corrigan:1978ce,Christ:1978jy,Corrigan:1978xi},
and 
\begin{eqnarray}
 G_A^{(0)}(x,y) = O(|x-y|^{-2}) \ %, |x-y| >>1 \ 
. \label{green2}
\end{eqnarray}
%In \cite{maeda_sako2},
%we see $| A^{(l)} | < O( |x|^{-3+\epsilon} )$ for arbitrary
%$\epsilon$ and $(l > 0)$ .
{}From (\ref{AsymptoticA}) and (\ref{3epsilon}) $ A^{(l)} = O( |x|^{-3+\epsilon} )$
we found that 
\begin{eqnarray}
[ \Delta_A \sum_{0\le k <n} \hbar^k G_A^{(k)} (x,y) ]^{(n)} = O(|x-y|^{-5}) \ .
\end{eqnarray}
Therefore,
\begin{eqnarray}
 G_A^{(n)} (x,y) = O( |x-y|^{-3} ) \ . \label{green3}
\end{eqnarray} 

%%%%%%%%%%%%%%%%%%%%%%%%%%%%%%%%%%%%%%%%%%%%%%%%%%%%%%%
%%%%%%%%%%%%%%%%%%%%%%%%%%%%%%%%%%%%%%%%%%%%%%%%%%%%%%%

\section{From Instantons to The ADHM Equations}
\label{InstantonADHM}

In this section we derive the ADHM equations from a
noncommutative instanton.

We let $\star_x $ denote $\star$ with 
respect to the variable $x=(x_1 , \dots , x_4)$.
Let $\bar{\psi}_i\ (i=1,\dots ,k) $ be orthonormal zero modes
of $\displaystyle \Diracb \star$ and set matrix
$\bar{\psi} = (\bar{\psi}_i )$ as
in Section  \ref{section_Index}.
%%%%%%%%%%%%%%% rev10
The concept of completeness in the Hilbert spaces 
is extended to the one in formal expansion spaces, and
we obtain the following identity for arbitrary functions $f(x) , g(y)$.
\begin{eqnarray}
&&\int_{{\mathbb R}^4} \!\!\!\!\! d^4 x 
\int_{{\mathbb R}^4} \!\!\!\!\! d^4 y 
\ f(x) \star_x \bar{\psi} (x) \bar{\psi}^{\dagger} (y) \star_y g(y) 
\label{compl} \\
&&= 
\int_{{\mathbb R}^4} \!\!\!\!\! d^4 x 
\int_{{\mathbb R}^4} \!\!\!\!\! d^4 y 
\big\{ f(x) \star_x \delta (x-y) \star_y g(y) \nonumber \\
&& \ \ \ 
- f(x) \star_x \Dirac \star_x G_A (x,y) \star_y \overleftarrow{\Diracb} 
 \star_y g(y) \big\}, \nonumber
\end{eqnarray}
The proof for (\ref{compl}) is given in Appendix \ref{spinor_completeness}.
%%%%%%%%%%%%%%% rev10
%Note that $\displaystyle \int_{{\mathbb R}^4} d^4 y 
%f(x) \delta(x-y) g(y) = \int_{{\mathbb R}^4} d^4 y 
%f(x) \delta(x-y) g(y)
For consistency, we impose the
commutation
\begin{eqnarray}
[x^{\mu} , y^{\nu} ]_\star = \left\{
\begin{array}{cc}
i \theta^{\mu \nu} , & (x=y), \\
0 ,& (x \neq y ).
\end{array}
\right. .
\end{eqnarray}

In the following derivation of the ADHM equations,
we use the completeness condition and 
the asymptotic behavior of the zero modes
of the $\Diracb \star$ given by Theorem \ref{lemma_A}.

We first define $T^{\mu}$ by
\begin{eqnarray}
T^{\mu} &:=& \int_{{\mathbb R}^4} d^4 x 
\frac{1}{2} \left( 
x^{\mu} \star \bar{\psi }^{\dagger} \star \bar{\psi} + 
\bar{\psi }^{\dagger} \star \bar{\psi}
\star x^{\mu}
\right) \label{def_T}\\
&=& \int_{{\mathbb R}^4} d^4 x 
( x^{\mu} \star \bar{\psi }^{\dagger} \star \bar{\psi} )
= \int_{{\mathbb R}^4} d^4 x 
( \bar{\psi }^{\dagger} \star \bar{\psi} \star x^{\mu} ).
\nonumber
\end{eqnarray}
Here we use 
$\displaystyle \int_{{\mathbb R}^4} d^4 x
\partial_{\mu}( \bar{\psi }^{\dagger} \star \bar{\psi } ) =0$
in the second and third equalities in (\ref{def_T}),
%Note that $\displaystyle \int_{{\mathbb R}^4} d^4 x
%\partial_{\mu}( \bar{\psi } \star \psi ) =0$ is given by
which follows from 
$\bar{\psi }= O(|x|^{-3})$ (see
Theorem \ref{lemma_A} ).
Then,
\begin{eqnarray}
T^{\mu} T^{\nu} 
= \int_{{\mathbb R}^4} d^4 x 
\int_{{\mathbb R}^4} d^4 y 
( x^{\mu} \star_x  \bar{\psi }^{\dagger}(x) \star_x \bar{\psi} (x) ) (
\bar{\psi }^{\dagger}(y) \star_y \bar{\psi} (y) \star_y y^{\nu} )  \label{TT0}
\end{eqnarray}
Using (\ref{compl}) and integration by parts,  (\ref{TT0}) becomes
\begin{eqnarray}
T^{\mu} T^{\nu} &=& \int_{{\mathbb R}^4} d^4 x \
x^{\mu} \star \bar{\psi}^{\dagger} \star \bar{\psi} \star x^{\nu} \nonumber \\
&&+ \int_{S^3} d S_x^{\rho} \int_{{\mathbb R}^4} d^4 y 
(x^{\mu} \star_x \bar{\psi}^{\dagger}(x) \sigma_{\rho})\star_x 
G_A(x,y) \star_y \overleftarrow{\Diracb} \star_y
( \bar{\psi} (y) \star_y y^{\nu} )  \nonumber \\
&&- \int_{{\mathbb R}^4} d^4 x 
\int_{{\mathbb R}^4} d^4 y 
( \bar{\psi}^{\dagger} (x) \sigma^{\mu})\star_x 
G_A(x,y)\star_y \overleftarrow{\Diracb} \star_y
( \bar{\psi} (y) \star_y y^{\nu} ) ,\nonumber 
\end{eqnarray}
where $ d S_x^{\mu}=|x|^2 x^{\mu} d\Omega $ and $d \Omega$ is the
solid angle.
The first term
is deformed as follows.
\begin{eqnarray}
&&\displaystyle \int_{{\mathbb R}^4} d^4 x \
x^{\mu} \star \bar{\psi}^{\dagger} \star \bar{\psi} \star x^{\nu}
 \nonumber \\
&&=\int_{{\mathbb R}^4} d^4 x \big(
 \bar{\psi}^{\dagger} \star \bar{\psi} \star x^{\nu}\star x^{\mu} 
+ [x^{\mu} , \bar{\psi}^{\dagger} \star \bar{\psi} ]_\star
\star x^{\nu} +
\bar{\psi}^{\dagger} \star \bar{\psi}\star [x^{\mu} , x^{\nu} ]_\star
\big)
\nonumber \\
&&=  \int_{{\mathbb R}^4} d^4 x \big(
 \bar{\psi}^{\dagger} \star \bar{\psi} \star x^{\nu}\star x^{\mu} 
+ i \theta^{\mu \rho}\partial_{\rho} (\bar{\psi}^{\dagger} \star \bar{\psi})
\star x^{\nu} + 
i \theta^{\mu \nu} \bar{\psi}^{\dagger} \star \bar{\psi} \big)
\nonumber \\
&&=  \int_{{\mathbb R}^4} d^4 x ~
 \bar{\psi}^{\dagger} \star \bar{\psi} \star x^{\nu}\star x^{\mu} .
\end{eqnarray}
Here $\bar{\psi}=O(|x|^{-3})$ is used in the third equality.
By integration by parts again, we get
\begin{eqnarray}
&&T^{\mu} T^{\nu}=
\nonumber \\
&&
 \int_{{\mathbb R}^4} d^4 x ~
 \bar{\psi}^{\dagger} \star \bar{\psi} \star x^{\nu}\star x^{\mu}  
\label{TT1} \\
&& + \int_{S^3} d S_x^{\rho} \int_{S^3} d S_y^{\tau} 
(x^{\mu} \star_x \bar{\psi}^{\dagger}(x) \sigma_{\rho}) \star_x 
G_A(x,y) \star_y 
(\bar{\sigma}_{\tau} \bar{\psi} (y) \star_y y^{\nu} )  \label{TT2} \\
&& - \int_{S^3} d S_x^{\rho} \int_{{\mathbb R}^4} d^4 y 
(x^{\mu} \star_x \bar{\psi}^{\dagger} (x) \sigma_{\rho} )\star_x 
G_A(x,y) \star_y
( \bar{\sigma}^{\nu} \bar{\psi} (y)  )  \label{TT3} \\
&& - \int_{{\mathbb R}^4} d^4 x 
\int_{S^3} d S_y^{\tau} 
( \bar{\psi}^{\dagger} (x) \sigma^{\mu}) \star_x 
G_A(x,y) \star_y
(\bar{\sigma}^{\tau}  \bar{\psi} (y) \star_y y^{\nu} )  \label{TT4}\\
&& + \int_{{\mathbb R}^4} d^4 x \int_{{\mathbb R}^4} d^4 y 
(\bar{\psi}^{\dagger} (x) \sigma^{\mu} ) \star_x 
G_A(x,y) \star_y
(\bar{\sigma}^{\nu}  \bar{\psi} (y) ) . \label{TT5}
\end{eqnarray}

(\ref{TT2}) and (\ref{TT4}) vanish when $R_y \rightarrow \infty$,
where $R_y$ is a radius of $S_y^3$.
(\ref{TT5}) will vanish on the selfdual projection
$[T^{\mu} , T^{\nu} ]^+ := 
P^{\mu \nu , \rho \tau} [T_{\rho} , T_{\tau} ] $ 
(see (\ref{SDProjection})),
because 
$\sigma^{\mu} \bar{\sigma}^{\nu} - \sigma^{\nu} \bar{\sigma}^{\mu}$
is anti-selfdual with respect to the $\mu , \nu$.
Thus only (\ref{TT1}) and (\ref{TT3}) remain.

We introduce an asymptotically parallel section
$g^{-1} S$ of ${\cal S}^+ \otimes E$ by
\begin{eqnarray}
\tilde{\psi} = - \frac{g^{-1} S x^{\dagger}}{|x|^4} + O(|x|^{-4}),
\label{asym_psi}
\end{eqnarray}
where $S$ is a constant matrix, $x^{\dagger} := \bar{\sigma}_{\mu} x^{\mu}$,
$\tilde{\psi} := {}^t \bar{\psi} \sigma_2$, and 
${}^t$ means transposing the spinor indices. (See also Appendix
\ref{appendix}.)
Recall that $A$ has asymptotic behavior given by
(\ref{AsymptoticB}), and
note that $D_{\mu}\star g^{-1} \rightarrow 0$ as $r \rightarrow \infty$.
Using these facts and $\Diracb \star \bar{\psi} =0 $, we can prove 
that $\tilde{\psi}$ has the expression
(\ref{asym_psi}) by direct calculations similar to the commutative case.
Note that $\tilde{\psi}$ and $\bar{\psi}$ 
have one-to-one correspondence and $D_{\mu} \star \tilde{\psi} \sigma^{\mu} =0$
iff $\Diracb \star \bar{\psi}= 0$ (see Appendix \ref{appendix}).
 
Let us introduce $\chi$ by
\begin{eqnarray}
\chi (x) := 4 \pi \int_{{\mathbb R}^4} d^4 y~ G_A(x,y) \tilde{\psi} (y)
=4 \pi \int_{{\mathbb R}^4} d^4 y~ G_A(x,y) \star_y \tilde{\psi} (y) .
\label{def_chi}
\end{eqnarray}

\begin{lem}
$\chi$ is given asymptotically by
\begin{eqnarray}
\chi = - \frac{g^{\dagger} S x^{\dagger}}{|x|^2} + O(|x|^{-2}) \ .
\end{eqnarray}
\end{lem}

\noindent
\begin{pf}   %Note page chi-4 ,chi-5
%Let introduce $\tilde{\psi}$ by
%\[
%\tilde{\psi} = {}^t \bar{\psi} \sigma_2 
%=\frac{1}{\pi}V^{\dagger} \star C \star f ,
%\]
%where ${}^t$ means transposing spinor suffixes.
%{}From the lemma \ref{lemma_A},
%the surviving terms in noncommutative deformation of $\tilde{\psi}$
Consider
\begin{eqnarray}
\Delta_A \star (|x|^2 \tilde{\psi} )
= 8 \tilde{\psi} + 4 x^{\mu} (D_{\mu} \star \tilde{\psi})
+ |x|^2 (D^{\mu} \star D_{\mu} \star \tilde{\psi} ) + O(|x|^{-4}) .
\label{lem2_1}
\end{eqnarray}
Here 
\begin{eqnarray}
x^{\mu}(D_{\mu} \star O(|x|^{-4}) ) &=& O(|x|^{-4}) \ ,
\label{lem2_2}
\end{eqnarray}
and
\begin{eqnarray}
x^{\mu}(D_{\mu} \star \frac{g^{\dagger} S x^{\dagger}}{|x|^4} ) 
&=& -3 \frac{g^{\dagger} S x^{\dagger}}{|x|^4} + O(|x|^{-5}) .
\label{lem2_3}
\end{eqnarray}
Using (\ref{lem2_2}) and (\ref{lem2_3}), we have
\begin{eqnarray}
x^{\mu} D_{\mu}\star \tilde{\psi} = -3 \tilde{\psi} + O(|x|^{-4}) .
\label{lem2_4}
\end{eqnarray}

Note that
\begin{eqnarray}
D^{\mu} \star D_{\mu} \star O(|x|^{-4}) = O(|x|^{-6})
\label{lem2_5}
\end{eqnarray}
and 
\begin{eqnarray}
D^{\mu} \star D_{\mu} \star \frac{g^{\dagger} S x^{\dagger}}{|x|^4} =
O(|x|^{-6}) . \label{lem2_6}
\end{eqnarray}
Thus, we get
\begin{eqnarray}
D^{\mu} \star D_{\mu} \star\tilde{\psi}  = O(|x|^{-6}) .
\label{lem2_7}
\end{eqnarray}
{}From (\ref{lem2_4}) and (\ref{lem2_7}),
\begin{eqnarray}
\Delta_A \star (|x|^2 \tilde{\psi} ) = -4 \tilde{\psi} + O(|x|^{-4}) 
\end{eqnarray}
Applying the Green's function and using 
(\ref{green2}) and (\ref{green3}), 
we get the desired result.
\mbox{}\hfill$\square $
\end{pf}

Note that 
\begin{eqnarray}
\Dirac^2 \star \chi= -4\pi \tilde{\psi} . \nonumber 
\end{eqnarray}
By this relation and the asymptotic behaviors of $\chi$ and $\bar{\psi}$, 
(\ref{TT3}) becomes
\begin{eqnarray}
\frac{1}{8} tr (S^{\dagger} S \bar{\sigma}^{\mu} \sigma^{\nu}),
\end{eqnarray}
where the trace $tr$ is taken with respect to the 
spinor indices.

In the $[T^{\mu} , T^{\nu}]^+$ combination,
(\ref{TT1}) becomes $-i\theta^{\mu \nu +}=
-i P^{\mu \nu , \rho \tau} \theta^{\rho \tau}$.

Then from (\ref{TT1})-(\ref{TT5}) and the definition of
$[T^{\mu} , T^{\nu}]^+$, we obtain the following theorem.
%%%%%%%%%%%%%%%
\begin{thm}
Let $A^{\mu}$ be a SNCD instanton, and
$\bar{\psi}$ be the zero mode of $\Diracb \star$ determined by
$A^{\mu}$ as in Section \ref{section_Index}.
Let $T^{\mu}, S$ be constant matrices defined by (\ref{def_T}) and 
(\ref{asym_psi}), respectively.
Then, they satisfy the ADHM equations:
\begin{eqnarray}
[T^{\mu} , T^{\nu} ]^+ = \frac{1}{2} tr (S^{\dagger} S \bar{\sigma}^{\mu \nu})-
i\theta^{\mu \nu +} I_{k\times k} . \label{ADHM}
\end{eqnarray}
\end{thm}

These ADHM equations are the same as those given
by Nekrasov and Schwarz \cite{NCinstanton}.\\

In \cite{NCinstanton}, it is shown that
instantons can be constructed from ADHM data satisfying (\ref{ADHM}).
The spinor zero modes of the Dirac operator in a background 
of noncommutative ADHM instantons are studied, 
and the index of the Dirac operator is given in
\cite{Kim:2002qm}.
The question of whether there is a one-to-one correspondence between 
ADHM data and instantons is answered affirmatively.
%%%%%%%%%%%%%%%
\begin{thm}
There is a one-to-one correspondence between 
ADHM data satisfying (\ref{ADHM}) and SNCD instantons
in noncommutative ${\mathbb R}^4$.
\end{thm}
The proof for this theorem is given in Appendix
\ref{appendix}.\\

It may be useful to note the relation between Theorem \ref{lemma_A}
and the term $S$ in the ADHM data.
$S$ is given as the coefficient
of the $O(|x|^{-3})$ term in $\bar{\psi}$, and 
Theorem \ref{lemma_A} implies that the 
$O(|x|^{-3})$ term is a zero mode of 
$\Dirac^{(0)}$.
For example, when we consider $k=1$,
there is only one zero mode $\bar{\psi}$.
One might think that
the $O(|x|^{-3})$ term in each $\bar{\psi}^{(n)}$
is proportional to $\bar{\psi}^{(0)}$, and 
$S^{(n)}$ is also proportional to $S^{(0)}$, but this is 
not true in general, due to
gauge symmetries and global symmetries.
$g^{\dagger}$ can also be expanded as a power series in $\hbar$
(see Appendix \ref{appendixB}).
For example, $\tilde{\psi}^{(1)}$ is given by
\begin{eqnarray}
\tilde{\psi}^{(1)} = - 
\frac{\{ (g^{\dagger})^{(0)} S^{(1)} + (g^{\dagger})^{(1)} S^{(0)}\}x^{\dagger}}{|x|^4} 
+ O(|x|^{-4}) .
\end{eqnarray}
As a result of this twisting by $(g^{\dagger})^{(1)}$,
$S^{(1)}$ is not proportional to $S^{(0)}$ in general,
and so $tr (S^{\dagger} S \bar{\sigma}^{\mu} \sigma^{\nu})$
is not proportional to 
$tr (S^{(0)\dagger} S^{(0)} \bar{\sigma}^{\mu} \sigma^{\nu})$.
In fact, taking the trace of (\ref{ADHM}) shows that 
$Tr \{ tr (S^{\dagger} S \bar{\sigma}^{\mu} \sigma^{\nu}) \}$
is deformed by the noncommutative
parameter from $0$ to $i k \theta^{\mu \nu +}$, 
where trace $Tr$ is taken with respect to the $k\times k$ matrix indices.

%%%%%%%%%%%%%%%%%%%%%%%%%%%%%%%%%%%%%%%%%%%%%%%%%%%%%%%%%%%%%
\section{Example}
In this section, we compute a simple example of a 
noncommutative instanton that is deformed smoothly from
a commutative one.
The notation used in this section is given 
in Appendix \ref{notation}.\\

We start from a $U(2)$ BPST instanton in commutative $\mathbb{R}^4$
with the instanton number $k=-1$ \cite{Belavin:1975fg}.
Its ADHM data is given by
\begin{eqnarray}
T^{\mu}= b^{\mu} ,\ S=\left(
\begin{array}{cc}
\rho \ & 0\ \\
0\  & \rho\ 
\end{array}
\right) ,\  \rho , b^{\mu} \in \mathbb{R} .
\label{ADHM_ex_com_BPS}
\end{eqnarray}
The ADHM data (\ref{ADHM_ex_com_BPS}) satisfies 
\begin{eqnarray}
[T^{\mu} , T^{\nu} ]^+ = \frac{1}{2} tr (S^{\dagger} S \bar{\sigma}^{\mu \nu}) .
\nonumber
\end{eqnarray}
We deform the ADHM equations to the (\ref{ADHM}).
For simplicity, we set 
\begin{eqnarray}
\theta^{12}=-\theta^{21}=\hbar , \ \ 
\theta^{\mu \nu}=0  \ ( (\mu, \nu) \neq (1,2) , (2,1) )
\end{eqnarray}
in this section.
Then the ADHM data satisfying (\ref{ADHM}) deforms to
\begin{eqnarray} \label{ADHM_example}
T^{\mu}= b^{\mu} ,\ S=\left(
\begin{array}{cc}
\sqrt{\rho^2 + \hbar} & 0 \\
0 & \sqrt{\rho^2 - \hbar}
\end{array}
\right) .  
%\rho , b^{\mu} \in \mathbb{R} .
\end{eqnarray}
Note that the data (\ref{ADHM_example}) connects
 smoothly to (\ref{ADHM_ex_com_BPS})
in the commutative limit, 
and the noncommutative deformation of the ADHM data is not unique.
By setting $y^{\mu}= x^{\mu} -b^{\mu}$,
the solution of $\nabla^{\dagger} \star \tilde{V} =O$ is given by
\begin{eqnarray}
\tilde{V} =(\tilde{V}_1 \tilde{V}_2)= \left(
\begin{array}{c}
\bar{\sigma}_{\mu} y^{\mu}\\
M
\end{array}
\right) , 
\end{eqnarray}
where $\tilde{V}_i$ is a 4-vector and
\begin{eqnarray}
M&:=&
-(\bar{\sigma}_{\mu} y^{\mu})^{-1}_{\star}
\star
\left(
\begin{array}{cc}
\sqrt{\rho^2 + \hbar} & 0 \\
0 & \sqrt{\rho^2 - \hbar}
\end{array}
\right)
\star
(\bar{\sigma}^{\nu} y_{\nu}) .
\end{eqnarray}
Here $(\bar{\sigma}_{\mu} y^{\mu})^{-1}_{\star}$
is defined by $(\bar{\sigma}_{\mu} y^{\mu})^{-1}_{\star} 
\star (\bar{\sigma}_{\mu} y^{\mu})=I_{2 \times 2}$.
Expanding $M$ as $\displaystyle M=\sum_{k=0}^{\infty}
M^{(k)} \hbar^k$, we have
\begin{eqnarray}
&&M^{(0)}= -\rho I_{2 \times 2} , \nonumber \\
&&M^{(1)}={\cal M} +O(|x|^{-2}), \ \ 
{\cal M}=-\frac{1}{2\rho |y|^2} y 
\left(
\begin{array}{cc}
1 & 0 \\
0 & -1
\end{array}
\right)y^{\dagger} ,
\end{eqnarray}
where $y:= y^{\mu} \sigma_{\mu} $ and 
$y^{\dagger}:= y^{\mu} \bar{\sigma}_{\mu} $.
We set the orthonormalization condition $V^{\dagger} \star V=I_{2\times2}$
for the solution of $\nabla^{\dagger} \star {V} =O$.
This normalization is not elementary because of the $\star$ product,
even if we use the Gram-Schmidt process, i.e.
\begin{eqnarray}
V_1 &:=& \tilde{V}_1 \star |\tilde{V}_1|^{-1}_{\star} , \nonumber \\
V_2^{\perp}  &:=& \tilde{V}_2 -V_1 \star (V_1^{\dagger} \star 
\tilde{V}_2 )  , \\
V_2 &:=& {V}_2^{\perp} \star |{V}_2^{\perp}|^{-1}_{\star} , \nonumber
\end{eqnarray}
where $|\tilde{V}_I|^{-1}_{\star}$ is defined by
$|\tilde{V}_I| \star |\tilde{V}_I|^{-1}_{\star} =1$.
The explicit expressions for $V_1$ and $V_2$ are given by
\begin{eqnarray}
V_1 &=&
\frac{1}{\sqrt{|y|^2+\rho^2}}
\left(
\begin{array}{c}
z_2 \\
-\bar{z}_1 \\
-\rho \\
0
\end{array}
\right)
+
\frac{\hbar}{2\rho |y|^2 \sqrt{|y|^2+\rho^2}}
\left(
\begin{array}{c}
0 \\
0 \\
|z_2|^2 -|z_1|^2 \\
2\bar{z}_1 z_2
\end{array}
\right) 
\nonumber \\
&&+O(\hbar^2) +O(|x|^{-2}) ,
\end{eqnarray}
\begin{eqnarray}
V_2 &=&
\frac{1}{\sqrt{|y|^2+\rho^2}}
\left(
\begin{array}{c}
z_1 \\
\bar{z}_2 \\
0 \\
-\rho
\end{array}
\right)
+
\frac{\hbar}{2\rho |y|^2 \sqrt{|y|^2+\rho^2}}
\left(
\begin{array}{c}
0 \\
0 \\
2\bar{z}_2 \bar{z}_1 \\
|z_1|^2 -|z_2|^2
\end{array}
\right)  \nonumber \\
&&+O(\hbar^2) +O(|x|^{-2}) ,
\end{eqnarray}
where $z_1 =y_2 + iy_1$ and $z_2 =y_4 +iy_3$.
Finally, we obtain
\begin{eqnarray}
{V}=({V}_1 {V}_2)= 
\frac{1}{\sqrt{|y|^2+\rho^2}}
\left(
\begin{array}{c}
y^{\dagger} \\
M^{(0)}+ \hbar M^{(1)}
\end{array}
\right) + O(\hbar^2) +O(|x|^{-2}) . \label{V_example}
\end{eqnarray}
For this $V$,
the SNCD instanton is given
by
\begin{eqnarray}
&&A_{\mu} = V^{\dagger} \star \partial_{\mu} V 
\nonumber \\
&&= A^{(0)}_{\mu} +
\frac{\hbar}{\sqrt{|y|^2+\rho^2}} \left(
\begin{array}{cc}
-\rho & 0 \\
0 & \rho
\end{array}
\right)
\partial_{\mu} \Big\{
\frac{1}{2\rho |y|^2 \sqrt{|y|^2+\rho^2}}
\left(
\begin{array}{cc}
|z_2|^2- |z_1|^2 & 2\bar{z}_2 z_1 \\
2\bar{z}_1 z_2 & |z_1|^2- |z_2|^2
\end{array}
\right) 
\Big\}
\nonumber \\
&&+\frac{\hbar}{2\rho |y|^2 \sqrt{|y|^2+\rho^2}}
\left(
\begin{array}{cc}
|z_2|^2- |z_1|^2 & 2\bar{z}_2 z_1 \\
2\bar{z}_1 z_2 & |z_1|^2- |z_2|^2
\end{array}
\right) 
\left(
\begin{array}{cc}
-\rho & 0 \\
0 & \rho
\end{array}
\right)
\partial_{\mu}
\frac{1}{\sqrt{|y|^2+\rho^2}}
\nonumber \\
&& +O(\hbar^2)+O(|x|^{-4}) , \label{NCBPST}
\end{eqnarray}
where $A^{(0)}_{\mu}$ is a commutative
instanton from \cite{Belavin:1975fg}:
\begin{eqnarray}
A^{(0)}_{\mu}=
\frac{y_{\mu}I_{2\times 2} +\sigma_{\mu} y^{\dagger}}{|y|^2+\rho^2} .
\end{eqnarray}
$A^{(1)}_{\mu}$, the term in proportional to $\hbar$
in (\ref{NCBPST}), is $O(|x|^{-3})$,
and this fact is consistent with 
(\ref{3epsilon}) (see also \cite{maeda_sako2}).
As we show in Appendix \ref{appendix_1_2}, 
the zero mode of $\Diracb \star$ is given as 
$\tilde{\psi}=
\frac{1}{\pi}(V^{\dagger} C) \star f $, where $C$ and $f$
are defined in Appendix \ref{notation}.
Substituting our ADHM data (\ref{ADHM_example}) and (\ref{V_example}),
we get
\begin{eqnarray}
\tilde{\psi}
&=& \tilde{\psi}^{(0)}+\hbar \tilde{\psi}^{(1)}+O(\hbar^2)
+O(|x|^{-4}) \nonumber \\
&=& \tilde{\psi}^{(0)}+
\frac{\hbar}{\pi (|y|^2+\rho^2)^{\frac{3}{2}} }
M^{(1)} +O(\hbar^2)
+O(|x|^{-4}) 
\nonumber \\
&=& \tilde{\psi}^{(0)}+
\frac{\hbar}{\pi |y|^3 }
{\cal M} +O(\hbar^2)
+O(|x|^{-4}) ,
\label{example_zeromode}
\end{eqnarray}
where $\tilde{\psi}^{(0)}$ is a zero mode of $\Diracb^{(0)}$:
\begin{eqnarray}
\tilde{\psi}^{(0)}
=\frac{-\rho}{(|y|^2+\rho^2 )^{\frac{3}{2}}}
I_{2\times 2} .
\end{eqnarray}
{}By Theorem
\ref{lemma_A}, the $O(|x|^{-3})$ term in the (\ref{example_zeromode})
should satisfy $D_{\mu}^{(0)} \tilde{\psi}^{(1)} \sigma^{\mu}=0$,
as in this example, 
%the following equation should be satisfied,
\begin{eqnarray}
(\partial_{\mu} +A_{\mu}^{(0)})
\Big( \frac{1}{|y|^3} {\cal M} \Big) \sigma^{\mu}
=0 .
\end{eqnarray}
This equation is easily verified by direct calculation.
%This result is consistent with Theorem
%\ref{lemma_A}.

\section{Conclusion}
Noncommutative deformations of zero modes for the 
Dirac operator with SNCD instanton backgrounds, the Green's 
function for SNCD instantons, and the ADHM equations
are investigated. 
{}From Theorem \ref{lem_no_zeromode}, Theorem \ref{lemma_A} and the
solutions (\ref{diraczeromode}), we find that there are
no new zero modes of $\Dirac \star$ and $\Diracb \star$, so 
the (modified) index of the Dirac operator is unchanged under 
noncommutative deformation.
The asymptotic behavior of the zero mode of $\Diracb \star$
is computed. In particular, the $O(|x|^{-3})$ terms
in the zero modes of $\Diracb \star$ are obtained from the zero modes of
the Dirac operator in commutative space.
This result implies that the term $S$ in the ADHM data is constructed from a
linear combination of the corresponding $S$ in the ADHM data of 
commutative ${\mathbb R}^4$.
The Green's function with a background SNCD instanton
is also constructed recursively.
Using these zero modes and the Green's function,
we derive the noncommutative ADHM equations
and prove a one-to-one correspondence 
between the ADHM data and SNCD instantons.
One simple example is studied as confirmation of our results:
we deform $k=-1$ BPST instanton into the SNCD instanton
via the ADHM method.
Consistency checks are verified by comparing the 
term proportional to $\hbar$ in
%Then consistent results are obtained from the observation of
%the $\hbar$ proportional terms of 
the SNCD
instanton and the zero mode of the $\Diracb \star$.

Our method is based on the $\hbar$ expansion, which means
that noncommutative instantons whose commutative limits are
singular, such as $U(1)$ instantons, are not considered in this article.
The relation between the ADHM equations and a noncommutative
instanton with a singular commutative limit remains to
be investigated in a future work.\\

\noindent
{\bf Acknowledgement}\\
Y.M and A.S are supported by KAKENHI No.22654011
(Grant-in-Aid for Exploratory Research)
 and No.20740049
 (Grant-in-Aid for Young Scientists (B)), respectively.
We would like to thank Steven Rosenberg for his 
through reading, helpful suggestions and comments.
We would like to express our gratitude to Hiroshi Umetsu and Toshiya
Suzuki for fruitful discussions and suggestions 
over a long period of time.
We would like to thank Anca Tureanu, Masud Chaichian and Claus Montonen 
for many important suggestions and comments.
%We would like to thank Steven Rosenberg for his helpful suggestions.
We would like to give special thanks to Masashi Hamanaka 
for his through reading our paper and his comments,
and for access to his work, which enabled us to complete this article. 

%The authors appreciate for the referee's helpful comments.
%%%%%%%%%%%%%%%%%%%%%%%%%%%%%%%%%%%%%%%%%%%%%%%%%%%%%%%%%%%%%
%%%%%%%%%%%%%%%%%%%%%%%%%%%%%%%%%%%%%%%%%%%%%%%%%%%%%%%%%%%%%
\appendix

%%%%%%%%%%% rev11
\section{Derivation of (\ref{compl})} \label{spinor_completeness}
In this section, we derive (\ref{compl}).
The identity for commutative space is proved by
\cite{Corrigan:1983sv, Brown:1977eb}.
We extend the identity to our formal expansion space.
Let us introduce a $\hbar$-valued spinor propagator 
$P(x,y)= \sum_{i=0}^{\infty} P^{(i)}(x,y) \hbar^i$ in a
$k$-instanton background by
\begin{eqnarray}
\gamma^{\mu} D_{\mu} \star P(x,y) = \delta(x-y) - \sum_{n=1}^k 
\Psi_n (x) \Psi^{\dagger}_n(y) ,  \label{Dirac_Propagator}
\end{eqnarray}
where we use $\gamma$-matrices
\begin{eqnarray}
\gamma^{\mu} = \left(
\begin{array}{cc}
0 & \bar{\sigma}^{\mu} \\
\sigma^{\mu} & 0
\end{array}
\right) 
\end{eqnarray}
and zero modes $\Psi_n (x)$ of the Dirac operator $\gamma^{\mu} D_{\mu}$.
We expand (\ref{Dirac_Propagator}) in $\hbar$:
\begin{eqnarray}
\gamma^{\mu} D_{\mu}^{(0)} P^{(0)}(x,y) &=& \delta(x-y) - \sum_{i=1}^k 
\Psi_i^{(0)} (x) \Psi^{\dagger (0)}_i(y), \label{p(0)} \\
\gamma^{\mu} D_{\mu}^{(0)} P^{(1)}(x,y) &=& R^{(1)}(x,y), \nonumber \\
&\vdots & \nonumber \\
\gamma^{\mu} D_{\mu}^{(0)} P^{(n)}(x,y) &=& R^{(n)}(x,y), \label{eq_for_pn}\\
&\vdots & \nonumber
\end{eqnarray}
where 
\begin{eqnarray}
R^{(n)}(x,y) &:=& -\gamma^{\mu} A^{(n)}_{\mu} P^{(0)}(x,y)
- \!\!\! \sum_{(l~;k,m)\in I(n)} \!\!  \gamma^{\mu}
A_{\mu}^{(k)}(\overleftrightarrow{\Delta} )^l P^{(m)}(x,y)
\nonumber \\
&& -\sum_{i=1}^k  \sum_{j=1}^n 
\Psi_i^{(j)} (x) \Psi^{\dagger (n-j)}_i(y).
\end{eqnarray}
In the \cite{Brown:1977eb}, the existence of $P^{(0)}(x,y)$
satisfying (\ref{p(0)}) is shown.
We can solve recursively (\ref{eq_for_pn}) for $P^{(n)}(x,y) (n=1,2,3,\dots)$
by separating $P^{(n)}(x,y)$ into two parts by chirality 
and using the similar way of Section \ref{section_Index}.
Then we find that there exist 
$P(x,y)= \sum_{i=0}^{\infty} P^{(i)}(x,y) \hbar^i$
such that (\ref{Dirac_Propagator}).

Next, we derive (\ref{compl}).
Similar to \cite{Corrigan:1983sv}, we take the form
\begin{eqnarray}
P(x,y)=\left(
\begin{array}{cc}
0 & s(x,y) \\
\bar{s} (x,y) & 0
\end{array}
\right) ,
\end{eqnarray}
then $s(x,y)$ and $\bar{s}(x,y)$ satisfy
\begin{eqnarray}
\Diracb \star \bar{s}(x,y) &=& \delta(x-y), \\
\Dirac \star_x s(x,y) &=& \delta(x-y) - 
\sum_{i=1}^k \bar{\psi}_i(x) \bar{\psi}_i^{\dagger}(y) . \label{s_definition}
\end{eqnarray}
Here $\bar{\psi}_i(x)$ is a zero mode of $\Diracb \star$
given in Section \ref{section_Index}.
$\bar{s}(x,y)$ is obtained as
\begin{eqnarray}
\bar{s}(x,y) = \Dirac \star_x G_A(x,y). \label{sbar}
\end{eqnarray}
By multiplying $P^{\dagger}(z,x)$ from left side of (\ref{Dirac_Propagator}),
we obtain 
\begin{eqnarray}
-\bar{s}(z,y) + \int_{{\mathbb R}^4} d^4 x 
\sum_{i=1}^k \bar{\psi}_i(z) \bar{\psi}_i^{\dagger}(x) \star_x \bar{s}(x ,y)
= s^{\dagger}(z,y) .
\end{eqnarray}
Because of (\ref{sbar}),
\begin{eqnarray}
\mbox{}\!\!\!\!\!\!\!
\int_{{\mathbb R}^4} \!\! d^4 x 
\sum_{i=1}^k \bar{\psi}_i(z) \bar{\psi}_i^{\dagger}(x) \star_x \bar{s}(x ,y)
&=& - \int_{{\mathbb R}^4} \!\! d^4 x \sum_{i=1}^k 
\bar{\psi}_i(z) ( \Diracb \star_x \bar{\psi}_i(x))^{\dagger} ) 
\star_x G_A(x ,y) \nonumber \\
&=&0.\nonumber
\end{eqnarray}
Then the following relation is obtained:
\begin{eqnarray}
s(x,y) = -\bar{s}^{\dagger} (x,y). \label{s=-sbardagger}
\end{eqnarray}
{}From (\ref{s_definition}), (\ref{sbar}) and (\ref{s=-sbardagger}),
we obtain (\ref{compl}):
\begin{eqnarray}
\star_x \bar{\psi} (x) \bar{\psi}^{\dagger} (y) \star_y 
= 
\star_x \delta (x-y) \star_y 
-  \star_x \Dirac \star_x G_A (x,y) \star_y \overleftarrow{\Diracb} 
 \star_y . \nonumber
\end{eqnarray}
%%%%%%%%%% rev11

%%%%%%%%%%%%%%%%%%%%%%%%%%%%%%%%%%%%%%%%%%%%%%%%%%%%%%
\section{A One-to-One Correspondence between Instanton and ADHM Data}
\label{appendix}
In this Appendix, we prove a one-to-one correspondence 
between ADHM data and SNCD instanton solutions.
It is shown that
instantons can be constructed from ADHM data satisfying (\ref{ADHM})
in \cite{NCinstanton}.
The spinor zero modes of the Dirac operator in a background 
of noncommutative ADHM instantons are studied, 
and the index of the Dirac operator is given in
\cite{Kim:2002qm}.
In this paper, we show that the index of the Dirac operator 
does not depend on noncommutative parameters and the ADHM equations 
are constructed from SNCD instantons in this article.
The proof to show the one-to-one correspondence 
between ADHM data and SNCD instantons is completed if
we show the completeness and the uniqueness.
We will prove the completeness and the uniqueness 
in subsection \ref{appendix_1_2} and
\ref{appendix_1_3}, respectively.
In commutative ${\mathbb R}^4$, there is 
the same one-to-one correspondence
(see for example 
\cite{Corrigan:1983sv,Kronheimer_Nakajima,Charbonneau,hamanaka_soken}.
).
Many parts of the proofs for the completeness and the uniqueness 
are parallel to the commutative cases.

We use the asymptotic behavior of the SNCD instanton (\ref{3epsilon}) 
and the spinor zero modes (\ref{spinor_asym}) and other results 
derived from the decay conditions as needed throughout.

%up to the 
%zero modes of commutative Dirac operator $\Diracb^{(0)}$.
%To quote the zero modes, we fix the arbitrary constant $a_{n,i}^j$
%in the following sections.

\subsection{Notation for The ADHM Construction} \label{notation}
In this subsection, we set the notation for the ADHM construction. 
$(N+2k)\times 2k$ matrices $C$ and $\nabla$ are defined by
\[
C:= \left(
\begin{array}{c}
O_{N \times 2k } \\
I_{2k \times 2k}
\end{array}
\right)
, \ 
\nabla :=  
\left(
\begin{array}{c}
S \\
\sigma_{\mu} (x^{\mu} -T^{\mu} )
\end{array}
\right) .
\]
{}From this definition, we have
\begin{eqnarray}
\partial_{\mu} \nabla = \sigma_{\mu} C .
\end{eqnarray}
If $T^{\mu}$ and $S$ satisfy the ADHM equations
(\ref{ADHM}), we have the following:
\begin{eqnarray}
\nabla^{\dagger} \star \nabla =S^{\dagger} S + 
\bar{\sigma}^{\mu}
\sigma^{\nu}(x^{\mu}-T^{\mu}) \star ( x^{\nu}-T^{\nu})=
\left(
\begin{array}{cc}
\square & O_{k\times k} \\
O_{k\times k} & \square
\end{array}
\right) ,
\end{eqnarray}
where
\begin{eqnarray}
\square := \frac{1}{2} tr (D^{\dagger} D )
+2 T_{\mu} x^{\mu} +|x|^2 %+ \frac{i}{4} \theta^{\mu \nu} 
,  \nonumber \\
D=
\left(
\begin{array}{c}
-S \\
T
\end{array}
\right) .  \label{square}
\end{eqnarray}
Here $T=T^{\mu} \sigma_{\mu}$.

Let us introduce the $(N+2k)\times N$ matrix $V$ satisfying 
\begin{eqnarray}
\nabla^{\dagger} \star V &=& O , \\ 
V^{\dagger} \star V &=& I_{N \times N} , \\ 
V \star V^{\dagger} &=& I_{(N+2k) \times (N+2k)} - 
\nabla \star f \star \nabla^{\dagger} . 
\end{eqnarray}
Here
\begin{eqnarray}
f := \square^{-1}_\star \ ,
\end{eqnarray}
and we define 
$g^{-1}_\star $, the inverse of $g$, 
by $g \star g^{-1}_\star =1 $.

We obtain a noncommutative instanton solution as
\begin{eqnarray}
A_{\mu} = V^{\dagger} \star \partial_{\mu} V.
\end{eqnarray}

\subsection{Completeness: ADHM $\Rightarrow$ Instanton $\Rightarrow$ ADHM }
\label{appendix_1_2}
In this subsection, we start with ADHM data satisfying the 
ADHM equations (\ref{ADHM}) is given.

We can obtain an instanton from this ADHM data as in \cite{NCinstanton}.
We show that we can reproduce the ADHM data from the instanton.

%Away from this given ADHM data to obtain instanton is done by the same manner 
%of \cite{NCinstanton}.
%We reproduce the ADHM data from the instanton.

In this subsection, $\nabla$ 
%introduced in the previous subsection \ref{notation} 
%is treated as the one constructed of the give ADHM data.
is associated to the given ADHM data.

Let us introduce $\tilde{\psi}$ by
\[
\tilde{\psi} = {}^t \bar{\psi} \sigma_2 
,
\]
where the transpose ${}^t$ is with respect to spinor indices.
Using $\sigma_2 \sigma_{\mu} \sigma_2 = -{}^t\bar{\sigma}_{\mu}$,
we find that 
\begin{eqnarray}
\Diracb \star \bar{\psi} = 0 
\Leftrightarrow D_{\mu} \star \tilde{\psi} \sigma^{\mu}=0 .
\end{eqnarray}
Therefore, to show $\Diracb \star \bar{\psi} =0$,
it suffices to prove $D_{\mu} \star \tilde{\psi} \sigma^{\mu}=0$.

%The proof for (\ref{A8}) is as follows.

%%%%%%%%%%%%%%%%%
\begin{lem}
Set $\tilde{\psi}$ 
\begin{eqnarray}
\tilde{\psi} = {}^t \bar{\psi} \sigma_2 
=\frac{1}{\pi}V^{\dagger} \star (C  f) , \label{VCf}
\end{eqnarray}
where $V$ and $f$ are defined in the
previous subsection \ref{notation} with respect to 
the given ADHM data.
Then $\tilde{\psi}$ satisfies 
\begin{eqnarray}
D_{\mu} \star \tilde{\psi} \sigma^{\mu}=0. \label{A8}
\end{eqnarray}
\end{lem}
%%%%%%%%%%%%%%%%%%%%%
\begin{pf}
\begin{eqnarray}
\pi D_{\mu} \star \tilde{\psi} \sigma^{\mu}&=&
D_{\mu} \star (V^{\dagger} \star (C  f) ) \sigma^{\mu}
\nonumber \\
&=&(\partial_{\mu}  V^{\dagger} + 
(V^{\dagger} \star \partial_{\mu} V) \star V^{\dagger}) \star
( C\sigma^{\mu} f ) + V^{\dagger}\star C \sigma^{\mu} \star \partial_{\mu}f
\label{A_11} \\
&=& \partial_{\mu} V^{\dagger}
\star (1-V \star V^{\dagger}) \star (C \sigma^{\mu} f) 
-V^{\dagger} \star (C \sigma^{\mu} f )\star 
\partial_{\mu} (\nabla^{\dagger} \star \nabla )\star f ,
\nonumber
\end{eqnarray}
where we use $I= f \star (\nabla^{\dagger} \star \nabla )$ .
Using $1-V \star V^{\dagger} = \nabla \star f \star \nabla^{\dagger} $,
(\ref{A_11}) becomes
\[
\partial_{\mu} V^{\dagger}
\star (\nabla \star f \star \nabla^{\dagger})\star (C \sigma^{\mu} f) 
-V^{\dagger} \star (C \sigma^{\mu} f )\star 
\partial_{\mu} (\nabla^{\dagger} \star \nabla )\star f .
\]
Differentiating of $ V^{\dagger} \star \nabla =0 $,
we get 
$( \partial_{\mu} V^{\dagger} ) \star \nabla
=-V^{\dagger} \partial_{\mu} \nabla = 
-V^{\dagger} \sigma_{\mu} C$.
Therefore, 
\begin{eqnarray}
&& \pi D_{\mu} \tilde{\psi} \sigma^{\mu} = 
\label{A_12_+} \\
&& - V^{\dagger}
\star \left\{(C \sigma_{\mu} f \star \nabla^{\dagger})\star (C \sigma^{\mu} f) 
+4 (C  f )\star C^{\dagger} \nabla \star f
-2 (C  f )\star C^{\dagger} \nabla  \star f
\right\} . \nonumber
\end{eqnarray}
Since $\nabla^{\dagger} C = \bar{\sigma}_{\nu} (x^{\nu}-T^{\nu} )$ and $\sigma_{\mu} \bar{\sigma}_{\nu} \sigma^{\mu}=-2 \sigma_{\nu}$,
the first term in (\ref{A_12_+}) equals 
\begin{eqnarray}
- V^{\dagger}
\star (C \sigma_{\mu} f \star \nabla^{\dagger})\star (C \sigma^{\mu} f) 
&=& -V^{\dagger} \star C \sigma_{\mu} 
f \star \bar{\sigma}_{\nu} (x^{\nu}-T^{\nu} ) \sigma^{\mu} \star f
\nonumber \\
&=& 
2 V^{\dagger} \star C f \star C^{\dagger} \nabla \star f .
\end{eqnarray}
Then we obtain
\[
\pi D_{\mu} \star \tilde{\psi} \sigma^{\mu}=0 .
\]
\mbox{}\hfill$\square $
\end{pf}

Next, we show the following lemma.
\begin{lem} 
Let $\bar{\psi}$ be the zero mode of $\Diracb \star$
defined by (\ref{VCf}). Then, 
\begin{eqnarray}
\bar{\psi}^{\dagger} \star \bar{\psi}
=-\frac{1}{4\pi^2} \partial^2 f .
\end{eqnarray}
\end{lem}
\begin{pf}
\begin{eqnarray}
\bar{\psi}^{\dagger} \star \bar{\psi}
&=& tr ({}^t\bar{\psi}^{\dagger} \star {}^t\bar{\psi} )
\nonumber \\
&=& \frac{1}{\pi^2} tr 
(( f C^{\dagger} )\star V \star V^{\dagger} \star (C f) )
\nonumber \\
&=& \frac{1}{\pi^2} tr
(( f C^{\dagger} )\star (1_{N+2k} -\nabla \star f \star \nabla^{\dagger}) \star (C f) ) , \label{AIA3_A}
\end{eqnarray}
where $tr$ denote the trace with respect to spinor indices.
By definition,
\begin{eqnarray}
( f C^{\dagger} )\star (\nabla \star f \star \nabla^{\dagger}) 
\star (C f)
= f \star ((x^{\mu} -T^{\mu})\sigma_{\mu}) \star f
\star (\bar{\sigma}_{\nu} (x^{\nu}-T^{\nu})) \star f .
\nonumber \\
\label{AIA3_1} 
\end{eqnarray}
Differentiating $1= f \star \square $, 
where $\square$ is given by (\ref{square}),
we get
\begin{eqnarray}
f \star (\bar{\sigma}_{\nu} (x^{\nu}-T^{\nu})) \star f 
= -\frac{1}{2} \partial_{\nu} f \bar{\sigma}^{\nu} .
\label{AIA3_2} 
\end{eqnarray}
Using (\ref{AIA3_2}), (\ref{AIA3_1}) can be rewritten as
\begin{eqnarray}
( f C^{\dagger} )\star (\nabla \star f \star \nabla^{\dagger}) 
\star (C f)
=-\frac{1}{2} f \star ((x^{\mu} -T^{\mu})\sigma_{\mu}) \star
(\partial_{\nu} f \bar{\sigma}^{\nu}) .
\label{AIA3_1d}
\end{eqnarray}
Since
$(f C^{\dagger} ) \star (C f)= f\star f $ , 
$tr~ \sigma_{\mu} \bar{\sigma}_{\nu} = tr~ \delta_{\mu \nu} $,
and by (\ref{AIA3_1d}), (\ref{AIA3_A}) equals
\begin{eqnarray}
\frac{1}{\pi^2} tr (f\star f + \frac{1}{2} f \star (x^{\mu} -T^{\mu})
\star \partial_{\mu}f ) .
\label{AIA3_B}
\end{eqnarray}
{} From $\partial^2 ( \square \star f )=\partial^2 1 =0$ , 
we obtain the following identity:
\begin{eqnarray}
f\star f + \frac{1}{2} f\star (x^{\mu}-T^{\mu} ) \star \partial_{\mu}f
= -\frac{1}{8} \partial^2 f .
\label{AIA4_3}
\end{eqnarray}
Using this identity in (\ref{AIA3_B}),
we obtain 
\begin{eqnarray}
\bar{\psi}^{\dagger} \star \bar{\psi} = 
\frac{-1}{\pi^2} tr (\frac{1}{8} \partial^2 f ) = 
-\frac{1}{4\pi^2}  \partial^2 f . \label{AIA4_4}
\end{eqnarray}
\mbox{}\hfill$\square $
\end{pf}

For the next step, we show orthonormality.
\begin{lem}
If $\bar{\psi}$ is a $\Diracb \star$ zero mode given above, we have the orthonormal condition
\begin{eqnarray}
\int d^4 x~ \bar{\psi}^{\dagger} \bar{\psi} =1 .
\end{eqnarray}
\end{lem}
\begin{pf}
Define $|x|_{\star}^{-2}$ by
$|x|^2 \star |x|_{\star}^{-2} =1 $.
Explicitly, we have
\begin{eqnarray}
|x|_{\star}^{-2} &=& \frac{1}{|x|^2}
+ \hbar^2 \frac{1}{2} \theta^{\mu \nu}_0 {\theta_{0 \mu}}^{\tau}
\frac{1}{|x|^6} (-\delta_{\nu \tau} + 4 \frac{x_{\nu}x_{\tau}}{|x|^2})
+\cdots +\hbar^{n} O(\frac{1}{|x|^{2n+2}})+ \cdots
\nonumber \\
&=& \frac{1}{|x|^2} + O(|x|^{-6}) . 
\label{AIA6_1}
\end{eqnarray}
Then,
\begin{eqnarray}
f=\square^{-1}_\star =  |x|_{\star}^{-2} \star (1+
\frac{1}{2} tr (D^{\dagger} D ) |x|^{-2}_\star
-2 T_{\mu} x^{\mu} \star |x|^{-2}_\star  )^{-1}_{\star} .
\label{AIA6_2}
\end{eqnarray}
By (\ref{AIA4_4}), (\ref{AIA6_1}) and (\ref{AIA6_2}),
\begin{eqnarray}
\bar{\psi}^{\dagger} \bar{\psi}= \delta^4(x) + \partial^2 O(|x|^{-3}),
\end{eqnarray}
and therefore we obtain
$\int d^4 x~ \bar{\psi}^{\dagger} \bar{\psi} =1$.
\mbox{}\hfill$\square $
\end{pf}

Now we can show the completeness; that is to say,
we can show that the original ADHM data can be reproduced 
from the noncommutative ADHM instanton by the definition
(\ref{def_T}) and (\ref{asym_psi}).
%We make $T'$ with the instanton that is made with the ADHM data $T$.
\begin{thm}
Let $T$ and $S$ be ADHM data and let $A$ be a noncommutative
instanton constructed from the ADHM data.
Let $\bar{\psi}$ be the spinor zero mode of $\Diracb$ given above.
Define $T'$ and $S'$ by
\begin{eqnarray}
{T'}^{\mu}&=&\int d^4 x~ x^{\mu} \star \bar{\psi}^{\dagger} \star \bar{\psi} ,
\nonumber \\
\tilde{\psi} &=& - \frac{g^{-1} S' x^{\dagger}}{|x|^4} + O(|x|^{-4}).
\end{eqnarray}
Then
\[ T=T'\ \  \mbox{and}\ \ S=S' . \]
\end{thm} 
\begin{pf}
\begin{eqnarray}
{T'}^{\mu}&=&\int d^4 x~ x^{\mu} \star \bar{\psi}^{\dagger} \star \bar{\psi}
\nonumber \\
&=&-\frac{1}{4\pi^2} \int d^4 x~ x^{\mu} \star
\partial^2 f
\nonumber \\
&=&-\frac{1}{4\pi^2} \int dS^{3\nu}~ 
(x^{\mu} \partial_{\nu} -\delta^{\mu}_{\nu} ) \star f
\nonumber \\
&=&-\frac{1}{4\pi^2} \int dS^{3\nu}~ 
(x^{\mu} \partial_{\nu} -\delta^{\mu}_{\nu} )
\star
|x|_{\star}^{-2} \star (1+
\frac{1}{2} tr (D^{\dagger} D ) |x|^{-2}_\star
-2 T_{\rho} x^{\rho} \star |x|^{-2}_\star  )^{-1}_{\star} 
%
%\frac{1}{|x|^2} \star (1+
%\frac{1}{2} tr (D^{\dagger} D ) |x|^2
%+2 T_{\mu} x^{\mu}|x|^2  + |x|^2\frac{i}{4} \theta^{\mu \nu}? )^{-1}
\nonumber \\
&=&-\frac{1}{4\pi^2} \int dS^{3\nu}~ 
(x^{\mu} \partial_{\nu} -\delta^{\mu}_{\nu} )
\frac{1}{|x|^4}  (
-2 T_{\rho} x^{\rho}   ) 
\nonumber \\
&=& T^{\mu} .
\end{eqnarray}
The proof for $S=S'$ is given by a direct calculation
similar to the commutative case.
\mbox{}\hfill$\square $
\end{pf}

%%%%%%%%%%%%%%%%%%%
\subsection{Uniqueness : Instanton $\Rightarrow$ ADHM $\Rightarrow$
Instanton} \label{appendix_1_3}
In this subsection, we start with a some noncommutative 
instanton $A$.
Let $D^{\mu}$ be the covariant derivative associated with the given 
noncommutative instanton connection $A^{\mu}$.
We introduce $\tilde{\xi} , \tilde{\chi} $ by
\begin{eqnarray}
D^{\mu} \star D_{\mu} \star \tilde{\xi} = 0, \ 
D^{\mu} \star D_{\mu} \star \tilde{\chi} = - 4\pi \tilde{\psi}
\end{eqnarray}
with the boundary conditions as $|x| \rightarrow \infty$:
\begin{eqnarray}
\tilde{\xi} \rightarrow g^{\dagger}, \
\tilde{\chi} \rightarrow - \frac{g^{\dagger} S x^{\dagger}}{|x|^2} .
\end{eqnarray}
\begin{lem}
Let $V$ be
\begin{eqnarray}
V=\left(
\begin{array}{c}
\tilde{\xi}^{\dagger} \\ \tilde{\chi}^{\dagger}
\end{array}
\right) . \label{uniq_V}
\end{eqnarray}
Then
\begin{eqnarray}
\nabla^{\dagger} \star V =0 , \ \  V^{\dagger} \star V = I_{N\times N}   .
\end{eqnarray}
\end{lem}
\begin{pf} The identity $D^{\mu} \star D_{\mu} \star \tilde{\xi} = 0$
implies that 
$D_{\mu} \star \tilde{\xi}$
can be written as a linear combination of $\tilde{\psi} \sigma_{\mu}$:
\begin{eqnarray}
D_{\mu} \star \tilde{\xi} = \tilde{\psi} \sigma_{\mu} L , 
\end{eqnarray}
where $L$ is a $2k \times N$ matrix.
By orthonormality,
\begin{eqnarray}
4L&=& \int d^4 x~ \bar{\sigma}_{\mu} \tilde{\psi}^{\dagger} \star 
D_{\mu} \star \tilde{\xi}
\nonumber \\
&=& \int dS^{3\mu}~ \bar{\sigma}_{\mu} \tilde{\psi}^{\dagger} \star \tilde{\xi}
\nonumber \\
&=& \int d\Omega~ |x|^2 x^{\mu} \bar{\sigma}_{\mu} 
\big( \frac{-xS^{\dagger}g}{\pi |x|^4} \star g^{\dagger} \big) = -4\pi S^{\dagger} ,
\end{eqnarray}
which implies
\begin{eqnarray}
D_{\mu} \star \tilde{\xi}= - \pi \tilde{\psi} \sigma_{\mu} S^{\dagger} .
\label{Dxi}
\end{eqnarray}
A similar computation gives
\begin{eqnarray}
D_{\mu} \star \tilde{\chi}= \pi \tilde{\psi} \sigma_{\mu} T^{\dagger}
- \pi \tilde{\psi} \sigma_{\mu} \star x^{\dagger}. \label{Dchi}
\end{eqnarray}
{}From (\ref{Dxi}) and (\ref{Dchi}), we have
\begin{eqnarray}
D_{\mu} \star V^{\dagger} = - \pi \tilde{\psi} \star \sigma_{\mu} \nabla^{\dagger} .
\label{DVd}
\end{eqnarray} 
We show that $V^{\dagger} \star \nabla =0$. Note that
\begin{eqnarray}
D_{\mu} \star (V^{\dagger} \star \nabla)&=&
(D_{\mu}\star V^{\dagger} ) \star \nabla + V^{\dagger} \star \partial_{\mu} \nabla
\nonumber \\
&=& -\pi \tilde{\psi}\star (\sigma_{\mu} \nabla^{\dagger}) \star \nabla
+V^{\dagger} C \sigma_{\mu} ,
\end{eqnarray}
where we use (\ref{DVd}).
Then 
\begin{eqnarray}
D^{\mu} \star D_{\mu} (V^{\dagger} \star \nabla)&=& 
-\pi \tilde{\psi} \star \sigma_{\mu}
((\partial^{\mu} \nabla^{\dagger}) \star \nabla
+ \nabla^{\dagger}\star \partial^{\mu}\nabla)
+(D^{\mu} \star V^{\dagger} )C \sigma_{\mu}
\nonumber \\
&=&
-\pi \tilde{\psi} \star \sigma_{\mu}
( C^{\dagger}\sigma_{\mu} \nabla+ \nabla^{\dagger}\sigma_{\mu}C+
\nabla^{\dagger}  C \sigma_{\mu} )=0
\end{eqnarray}
As we saw in Section \ref{sectGreenFunction},
the Green's function of $D^{\mu} \star D_{\mu}= \Delta_A$
exists.
Therefore, we obtain
$(V^{\dagger} \star \nabla)=0$.

We now verify that $V^{\dagger} \star V=I_{N\times N}$. 
$V^{\dagger} \star V$ is a covariant constant, as
\begin{eqnarray}
D_{\mu} \star (V^{\dagger} \star V) &=&
(D_{\mu} \star V^{\dagger} ) \star V + V^{\dagger} \star (D_{\mu} \star V^{\dagger} )^{\dagger} 
\nonumber \\
&=& -\pi (\tilde{\psi} \star \sigma_{\mu} \nabla^{\dagger} \star V
+ V^{\dagger} \star \nabla \bar{\sigma}_{\mu} \star \tilde{\psi}^{\dagger} )=0 ,\end{eqnarray}
By its asymptotic behavior, 
$V^{\dagger} \star V \rightarrow g^{-1} \star g =I_{N\times N}$,
shows that $V^{\dagger} \star V=I_{N\times N}$.
\mbox{}\hfill$\square $
\end{pf}

Finally, we show the uniqueness of the noncommutative
ADHM instanton.
\begin{thm}
Let $A_{\mu}'$ be a noncommutative ADHM instanton 
constructed from $V$,
i.e. $A_{\mu}' = V^{\dagger} \star \partial_{\mu} V $ , where $V$ is defined in (\ref{uniq_V}).
Then, $A'$ is equal to $A$:
\begin{eqnarray}
A_{\mu}' =A_{\mu}.
\end{eqnarray}
\end{thm}
\begin{pf}
\begin{eqnarray}
A_{\mu}' &=& V^{\dagger} \star \partial_{\mu} V \nonumber \\
&=& V^{\dagger} \star (\partial_{\mu} V -V \star A_{\mu} ) 
+ V^{\dagger}\star V \star A_{\mu}  \nonumber \\
&=& V^{\dagger}\star (D_{\mu} \star V^{\dagger} )^{\dagger} +A_{\mu}
\nonumber \\
&=& -\pi V^{\dagger} \star \nabla \star \bar{\sigma}_{\mu} \tilde{\psi}^{\dagger} 
+A_{\mu} =A_{\mu} .
\end{eqnarray}
\mbox{}\hfill$\square $
\end{pf}
%%%%%%%%%%%%%%%%%%%%%%%%%%%%%%%
%%%%%%%%%%%%%%%%%%%%%%%%%%%%%%%
\section{Gauge Group Elements}
\label{appendixB}
In this Appendix, we study the conditions forced by the choice of the
$U(N)$ gauge group.
If $g \in G $ then $g^{\dagger} \star g = I_{N\times N}$.
By expanding $g$ as
$\displaystyle g= \sum_{i=0}^{\infty} g^{(i)} \hbar^i$,
each term in the equation $g^{\dagger} \star g = I_{N\times N}$ is 
given by
\begin{eqnarray}
\hbar^0 &:& (g^{\dagger})^{(0)} g^{(0)}  =  I_{N\times N} 
\label{B1}
\\
\hbar^1 &:&  (g^{\dagger})^{(1)} g^{(0)}+
(g^{\dagger})^{(0)} g^{(1)}+ \frac{i}{2}\theta^{\mu \nu}
\partial_{\mu} (g^{\dagger})^{(0)} \partial_{\nu} g^{(0)}=0 
\label{B2}\\
& \vdots & \nonumber \\
\hbar^n &:& (g^{\dagger})^{(n)} g^{(0)}+
(g^{\dagger})^{(0)} g^{(n)}+
\sum_{(p;m,l)\in I(n)} \frac{1}{p~ !} \big( (g^{\dagger})^{(m)} 
(\overleftrightarrow{\Delta} )^p g^{(l)}  \big) =0 \label{B3}\\
&\vdots& \nonumber
\end{eqnarray}
(\ref{B1}) show that $g^{(0)}$ is an element of the
$U(N)$ gauge group in commutative space.
Let us introduce a $N\times N$ Hermitian matrix $\phi(x)$
by $g^{(0)}=\exp i \epsilon \phi$ with infinitesimal 
gauge parameter $\epsilon$.
By expanding $g^{(1)}$ as
\begin{eqnarray}
g^{(1)}= \sum_{k=0}^{\infty} \epsilon^{k} g_k^{(1)} 
=\sum_{k=0}^{\infty} \epsilon^{k} (H^{(1)}_k + A^{(1)}_k),
\end{eqnarray}
where $H^{(1)}_k$ and $A^{(1)}_k$ are Hermitian part and 
anti-Hermitian part of $g^{(1)}_k$, respectively,
(\ref{B2}) becomes
\begin{eqnarray}
\epsilon^0 &:& H^{(1)}_0 =0 \\
\epsilon^1 &:& H^{(1)}_1 = \frac{i}{2}\{ A^{(1)}_0 , \phi \}\\
\epsilon^2 &:& H^{(1)}_2=
\frac{-1}{2} \left\{
i\{A^{(1)}_0 , \phi^2 \} 
+i[H^{(1)}_1 , \phi ]
-i\{ A^{(1)}_1 , \phi \} + \frac{i}{2}\theta^{\mu \nu}
\partial_{\mu} \phi \partial_{\nu} \phi 
\right\} \\
\epsilon^3 &:& H^{(1)}_3=
\frac{-1}{2} \left\{
i\{A^{(1)}_0 , \phi^3 \} 
-\{H^{(1)}_1 , \phi^2\}
+[A^{(1)}_1 , \phi^2 ] + i [ H^{(1)}_2 , \phi ]
-i\{ A^{(1)}_2 , \phi \} 
\right\} \nonumber\\
& \vdots &  .\nonumber 
\end{eqnarray}
These conditions show that we
can chose $A^{(1)}_k$ freely,
and the choice of $A_k^{(1)}$ determines 
$H^{(1)}_k$.
For example, it is possible to choose
$g^{(1)}$ as a non-zero 
constant matrix in the limit as $|x| \rightarrow \infty$.
Therefore we can not ignore the
asymptotic effect of $g^{(1)}$ in the estimation
of the ADHM data as mentioned in Section \ref{InstantonADHM}.

%%%%%%%%%%%%%%%%%%%%%%

%%%%%%%%%%%%%%%%%%%%%%%%%%%%%%%%%%%%%%%%%%%%%%%%%%%%%%%%


\begin{thebibliography}{999}

\bibitem{maeda_sako2}
Y. Maeda, A. Sako,
``Noncommutative Deformation of Instantons,''
J.Geom. Phys. {\bf 58} , 1784 (2008)
{\tt arXiv:0805.3373}. 

%\cite{Bayen:1977ha}
\bibitem{Bayen:1977ha}
  F.~Bayen, M.~Flato, C.~Fronsdal, A.~Lichnerowicz and D.~Sternheimer,
  ``Deformation Theory And Quantization. 1. Deformations Of Symplectic
  Structures,''
  Annals Phys.\  {\bf 111} (1978) 61.\\
  %%CITATION = APNYA,111,61;%%
F.~Bayen, M.~Flato, C.~Fronsdal, A.~Lichnerowicz and D.~Sternheimer,
  ``Deformation Theory And Quantization. 2. Physical Applications,''
  Annals Phys.\  {\bf 111} (1978) 111.
  %%CITATION = APNYA,111,111;%%




\bibitem{NCinstanton}
 N.~Nekrasov and A.~S.~Schwarz,
``Instantons on noncommutative $R^4$ and (2,0) 
superconformal six  dimensional
theory,''
Commun.\ Math.\ Phys.\  {\bf 198}, 689 (1998)
{\tt hep-th/9802068}. 

\bibitem{ADHM}
  M.~F.~Atiyah, N.~J.~Hitchin, V.~G.~Drinfeld and Yu.~I.~Manin,
  ``Construction of instantons,''
  Phys.\ Lett.\  A {\bf 65}, 185 (1978).
  %%CITATION = PHLTA,A65,185;%%


\bibitem{NCinstlecture}
    K.~Y.~Kim, B.~H.~Lee and H.~S.~Yang,
  ``Comments on instantons on noncommutative ${\mathbb R}^4$,''
  J.\ Korean Phys.\ Soc.\  {\bf 41}, 290 (2002)
  {\tt hep-th/0003093}. \\
  %%CITATION = JKPSD,41,290;%%
%
    K.~Furuuchi,
  ``Equivalence of projections as gauge equivalence on noncommutative  space,''
  Commun.\ Math.\ Phys.\  {\bf 217}, 579 (2001)
  {\tt hep-th/0005199}.  \\
  %%CITATION = CMPHA,217,579;%%
%
    N.~A.~Nekrasov,
  ``Noncommutative instantons revisited,''
  Commun.\ Math.\ Phys.\  {\bf 241}, 143 (2003)
  {\tt hep-th/0010017}.\\
  %%CITATION = CMPHA,241,143;%%
%
    K.~Furuuchi,
  ``Dp-D(p+4) in noncommutative Yang-Mills,''
  JHEP {\bf 0103}, 033 (2001)
  {\tt hep-th/0010119}. \\
  %%CITATION = JHEPA,0103,033;%%
%
  N.~A.~Nekrasov,
  ``Trieste lectures on solitons in noncommutative gauge theories,''
  {\tt hep-th/0011095}. \\
  %%CITATION = HEP-TH/0011095;%%
%
  D.~H.~Correa, G.~S.~Lozano, E.~F.~Moreno and F.~A.~Schaposnik,
  ``Comments on the U(2) noncommutative instanton,''
  Phys.\ Lett.\  B {\bf 515}, 206 (2001)
  {\tt hep-th/0105085}. \\
  %%CITATION = PHLTA,B515,206;%%
%
    O.~Lechtenfeld and A.~D.~Popov,
  ``Noncommutative multi-solitons in 2+1 dimensions,''
  JHEP {\bf 0111}, 040 (2001)
  {\tt hep-th/0106213}. \\
  %%CITATION = JHEPA,0111,040;%%
%
  T.~Ishikawa, S.~I.~Kuroki and A.~Sako,
  ``Elongated U(1) instantons on noncommutative ${\mathbb R}^4$,''
  JHEP {\bf 0111}, 068 (2001)
  {\tt arXiv:hep-th/0109111}. \\
  %%CITATION = JHEPA,0111,068;%%
%
   S.~Parvizi,
  ``Non-commutative instantons and the information metric,''
  Mod.\ Phys.\ Lett.\  A {\bf 17}, 341 (2002)
  {\tt hep-th/0202025}. \\
  %%CITATION = MPLAE,A17,341;%%
%
  N.~A.~Nekrasov,
  ``Lectures on open strings, and noncommutative gauge fields,''
  {\tt hep-th/0203109}. \\
  %%CITATION = HEP-TH/0203109;%%
%
   Y.~Tian and C.~J.~Zhu,
  ``Instantons on general noncommutative ${\mathbb R}^4$,''
  Commun.\ Theor.\ Phys.\  {\bf 38}, 691 (2002)
  {\tt hep-th/0205110}.\\
  %%CITATION = CTPMD,38,691;%%
%
  D.~H.~Correa, E.~F.~Moreno and F.~A.~Schaposnik,
  ``Some noncommutative multi-instantons from vortices in curved space,''
  Phys.\ Lett.\  B {\bf 543}, 235 (2002)
  {\tt hep-th/0207180}. \\
  %%CITATION = PHLTA,B543,235;%%
%
  F.~Franco-Sollova and T.~A.~Ivanova,
  ``On noncommutative merons and instantons,''
  J.\ Phys.\ A  {\bf 36}, 4207 (2003)
  {\tt hep-th/0209153}. \\
  %%CITATION = JPAGB,A36,4207;%%
%
   Y.~Tian and C.~J.~Zhu,
  ``Comments on noncommutative ADHM construction,''
  Phys.\ Rev.\  D {\bf 67}, 045016 (2003)
  {\tt hep-th/0210163}. \\
  %%CITATION = PHRVA,D67,045016;%%
%
  M.~Hamanaka,
``Noncommutative solitons and D-branes,''
{\tt hep-th/0303256}.
  %%CITATION = HEP-TH/0303256;%%
%


\bibitem{CommLimi}
O.~Lechtenfeld and A.~D.~Popov,
``Noncommutative 't Hooft instantons,''
J. High Energy Phys. {\bf 03} (2002) 040 
{\tt hep-th/0109209}. \\ %%CITATION = JHEPA,0203,040;%%
%
Z.~Horvath, O.~Lechtenfeld and M.~Wolf,
``Non-commutative instantons via dressing and splitting approaches,'' 
J. High Energy Phys. 0212 (2002) 060  {\tt hep-th/0211041}.





\bibitem{sako2}
T. Ishikawa, S. Kuroki and A. Sako,
``Instanton number  on noncommutative ${\mathbb R}^4$,''
{\tt hep-th/0201196}.
``{Calculation of the Pontrjagin class for U(1)
instantons on noncommutative ${\mathbb R}^4$},''
JHEP {\bf 0208}, 028 (2002) .

\bibitem{sako3}
A. Sako, ``Instanton number of noncommutative U(N) Gauge Theory,''
JHEP {\bf 0304}, 023 (2003) 
{\tt hep-th/0209139 }.


\bibitem{Furuuchi1}
K. Furuuchi,
{``Instantons on noncommutative ${\mathbb R}^4$ and projection operators,''}
Prog. Theor. Phys. {\bf 103}, {1043}, ({2000})  {\tt hep-th/9912047}.


\bibitem{Furuuchi2}
K. Furuuchi,
{``Topological charge of U(1) instantons,''}
Prog. Theor. Phys. Suppl. {\bf 144} ,79, (2001) 
{\tt hep-th/0010006}.

\bibitem{Tian}
Y. Tian, C. Zhu and X. Song,
{``Topological charge of noncommutative ADHM instanton,''}
Mod.Phys.Lett. {\bf A18},1691,(2003) {\tt hep-th/0211225}.

\bibitem{maeda_sako}
Y. Maeda, A. Sako,
{``Are vortex numbers preserved?,''}
J.Geom. Phys. {\bf 58} , 967(2008)
{\tt math-ph/0612041}.



\bibitem{sako4}
A. Sako, {``Noncommutative Deformation of Instantons and Vortexes,''}
J.Geom.Symm.Phys. {\bf 14} , 85(2009).\\
A. ~Sako,
 {``Recent developments in instantons in noncommutative ${\mathbb R}^4$,''}
 Adv.\ Math.\ Phys.\  {\bf 2010}, 270694(2010) .
  %%CITATION = ADMPA,2010,270694;%%




%%%%%%%%%%%%%%%%%%%%%%%%%%%%%%
%%%%%%   Moyal
%%%%%%%%%%%%%%%%%%%%%%%%%%%%
\bibitem{Moyal}
J. E. Moyal,``Quantum mechanics as a statistical theory,'' 
Proc. Cambridge Phil.Soc. {\bf 45}, 99 (1949) .


%%%%%%%%%%%%%%%%%%%%%%%%%%%%%
%%%%%%%%%%%%%%%%%%%%%%
\bibitem{D-K}
S.K. Donaldson and P.B. Kronheimer, 
``The Geometry of Four-Manifolds,'' 
{\it Oxford Math. Monographs, Oxford Univ. Press, 1990 } .



%\bibitem{Freed}
%  D.~S.~Freed and K.~K.~Uhlenbeck,
%  ``Instantons and Four - Manifolds,''
%\href{http://www.slac.stanford.edu/spires/find/hep/www?irn=1341413}{SPIRES entr%y}
%{\it  New York, USA: Springer (1984) 232 P. (Mathematical Sciences Research Ins%titute Publications, 1)}
%
%%%%%%%%%%%%%%%%%%%%%%%%


%%%%%%%%%%%%

%\cite{Kim:2002qm}
\bibitem{Kim:2002qm}
  K.~Y.~Kim, B.~H.~Lee and H.~S.~Yang,
  %``Zero-modes and Atiyah-Singer index in noncommutative instantons,''
  Phys.\ Rev.\  D {\bf 66}, 025034 (2002)
   {\tt hep-th/0205010}.
  %%CITATION = PHRVA,D66,025034;%%




\bibitem{Bartnik}
R.~Bartnik, ``The mass of an asymptotically flat manifold,''
 Commun.\ Pure Appl.\ Math.\ {\bf 39}, 661 (1986). 

%%%%%%%%%%
\bibitem{Charbonneau}
B.~Charbonneau, 
``Analytic aspects of periodic instantons,''
Cambridge, MA, USA 
(MIT PhD thesis, supervised by Tomasz Mrowka.) (2004)

%%%%%%%%%%


\bibitem{Corrigan:1978ce}
  E.~Corrigan, D.~B.~Fairlie, S.~Templeton and P.~Goddard,
  ``A Green's function for the general selfdual gauge field,''
  Nucl.\ Phys.\  B {\bf 140}, 31 (1978).
  %%CITATION = NUPHA,B140,31;%%

\bibitem{Christ:1978jy}
  N.~H.~Christ, E.~J.~Weinberg and N.~K.~Stanton,
  ``General self-dual Yang-Mills solutions,''
  Phys.\ Rev.\  D {\bf 18}, 2013 (1978).
  %%CITATION = PHRVA,D18,2013;%%


\bibitem{Corrigan:1978xi}
  E.~Corrigan, P.~Goddard and S.~Templeton,
  ``Instanton Green's functions and tensor products,''
  Nucl.\ Phys.\  B {\bf 151}, 93 (1979).
  %%CITATION = NUPHA,B151,93;%%
  




%\cite{Belavin:1975fg}
\bibitem{Belavin:1975fg}
  A.~A.~Belavin, A.~M.~Polyakov, A.~S.~Shvarts and Yu.~S.~Tyupkin,
  ``Pseudoparticle solutions of the Yang-Mills equations,''
  Phys.\ Lett.\  B {\bf 59}, 85 (1975).
  %%CITATION = PHLTA,B59,85;%%



%\cite{Corrigan:1983sv}
\bibitem{Corrigan:1983sv}
  E.~Corrigan and P.~Goddard,
  ``Construction Of Instanton And Monopole Solutions And Reciprocity,''
  Annals Phys.\  {\bf 154}, 253 (1984).
  %%CITATION = APNYA,154,253;%%

%%%%%%%%%%%%%%%%%%%%% rev ref


%\cite{Brown:1977eb}
\bibitem{Brown:1977eb}
  L.~S.~Brown, R.~D.~Carlitz, D.~B.~Creamer and C.~k.~Lee,
  ``Propagation functions in pseudoparticle fields,''
  Phys.\ Rev.\  D {\bf 17} (1978) 1583.
  %%CITATION = PHRVA,D17,1583;%%
%%%%%%%%%%%%%%%%%%%%%


\bibitem{Kronheimer_Nakajima}
 P.~B.~Kronheimer, H.~Nakajima, 
``Yang-Mills instantons on ALE gravitational instantons,'' 
Math.\ Ann.,\ {\bf 288} , 263 %-307 
(1990).


\bibitem{hamanaka_soken}
M.~Hamanaka, 
Soryushiron Kenkyu (Japanese Magazine)
, {\bf 106-1}, 1 (2002)\\
 M.~Hamanaka,
  ``Noncommutative solitons and D-branes,''
(Tokyo Univ. PhD thesis, 2003)  {\tt hep-th/0303256}.



\end{thebibliography}
\end{document}